\documentclass[journal]{IEEEtran}
\pdfoutput=1 
\usepackage{amsmath,graphicx,url,times,dsfont,amssymb}
\usepackage{amsmath, amssymb, amsopn}
\usepackage{subfigure}
\usepackage{acronym}
\usepackage{balance}
\usepackage{caption}
\usepackage{color}
\usepackage[usenames,dvipsnames]{xcolor}
\usepackage{amsthm}
\usepackage{algorithmic}
\usepackage[linesnumbered,ruled]{algorithm2e}
\usepackage{cite}

\newcommand{\BL}[1]{\textcolor{black}{#1}} %Maroon
\newcommand{\BLT}[1]{\textcolor{black}{#1}} %Maroon
\newcommand{\BLS}[1]{\textcolor{black}{#1}} %Maroon

\acrodef{TDOA}{time difference of arrival}
\acrodef{STFT}{short-time Fourier transform}
\acrodef{TF}{transfer function}
\acrodef{RTF}{relative transfer function}
\acrodef{GSC}{generalized sidelobe canceler}
\acrodef{LCMV}{linearly constrained minimum-variance}
\acrodef{RCB}{robust Capon beamformer}
\acrodef{FBF}{fixed beamformer}
\acrodef{BM}{blocking matrix}
\acrodef{NC}{noise canceler}
\acrodef{TF-GSC}{transfer function-\ac{GSC}}
\acrodef{DD-GSC}{Data Driven-\ac{GSC}}
\acrodef{RKHS}{reproducing kernel Hilbert space}
\acrodef{VV-RKHS}{vector-valued reproducing kernel Hilbert space}
\acrodef{VVMR}{vector-valued manifold regularization}
\acrodef{DOA}{direction of arrival}
\acrodef{SNR}{signal to noise ratio}
\acrodef{PSD}{power spectral density}
\acrodef{WLS}{weighted least square}
\acrodef{AIR}{acoustic impulse response}
\acrodef{RTF}{relative transfer function}
\acrodef{ATF}{acoustic transfer function}
\acrodef{WGN}{white Gaussian noise}
\acrodef{CPSD}{cross \ac{PSD}}
\acrodef{GMM}{Gaussian Mixture Model}
\acrodef{GCC}{generalized cross-correlation}
\acrodef{RMSE}{root mean square error}
\acrodef{DRR}{direct to reverberant ratio}
\acrodef{ML}{maximum likelihood}
\acrodef{MRL}{Manifold Regularized Localization}
\acrodef{KNN}{k-nearest neighbours}
\acrodef{DD-KNN}{Diffusion Distance k-nearest neighbours}
\acrodef{SVD}{singular value decomposition}
\acrodef{SMR}{Sequential Manifold Regularization}
\acrodef{MRL}{Manifold Regularization for Localization}
\acrodef{DDS}{Diffusion Distance Search}
\acrodef{EM}{Expectation Maximization}
\acrodef{MAP}{maximum a posteriori probability}
\acrodef{MMSE}{minimum mean squared error}
\acrodef{MUSIC}{multiple signal classification}
\acrodef{ESPRIT}{estimation of signal parameters via rotational invariance}
\acrodef{AEC}{acoustic echo cancellation}
\acrodef{SRP}{steered response power}
\acrodef{SRP-PHAT}{\ac{SRP}-phase transform}
\acrodef{MMGP}{multiple-manifold Gaussian process}
\acrodef{aRTF}{aggregated \ac{RTF}}

\newcommand\numberthis{\addtocounter{equation}{1}\tag{\theequation}}

\begin{document}
%\ninept

\title{Semi-Supervised Source Localization on Multiple-Manifolds with Distributed Microphones}
\author{Bracha~Laufer-Goldshtein~\IEEEmembership{Student Member,~IEEE}, Ronen~Talmon,~\IEEEmembership{Member,~IEEE} and Sharon~Gannot,~\IEEEmembership{Senior Member,~IEEE}
\thanks{Bracha~Laufer-Goldshtein and Sharon Gannot are with the Faculty of Engineering, Bar-Ilan University,
	Ramat-Gan, 5290002, Israel (e-mail: bracha.laufer@biu.ac.il, Sharon.Gannot@biu.ac.il); Ronen Talmon is with the Viterbi Faculty of Electrical Engineering, The Technion-Israel Institute of Technology, Technion City, Haifa 3200003, Israel, (e-mail: ronen@ee.technion.ac.il).}}

\markboth{ IEEE/ACM TRANSACTIONS ON AUDIO, SPEECH, AND LANGUAGE PROCESSING}
{Laufer \MakeLowercase{\textit{et al.}}: Semi-Supervised Source Localization on Multiple-Manifolds with Distributed Microphones}

\maketitle

\begin{abstract}
\BLT{The problem of source localization with ad hoc microphone networks in noisy and reverberant enclosures, given a training set of prerecorded measurements, is addressed in this paper. The training set is assumed to consist of a limited number of labelled measurements, attached with corresponding positions, and a larger amount of unlabelled measurements from unknown locations. However, microphone calibration is not required. We use a Bayesian inference approach for estimating a function that maps measurement-based feature vectors to the corresponding positions. The central issue is how to combine the information provided by the different microphones in a unified statistical framework. To address this challenge, we model this function using a Gaussian process with a covariance function that encapsulates both the connections between pairs of microphones and the relations among the samples in the training set. The parameters of the process are estimated by optimizing a \ac{ML} criterion. \BLS{In addition, a recursive adaptation mechanism is derived where the new streaming measurements are used to update the model.} Performance is demonstrated for 2-D localization of both simulated data and real-life recordings in a variety of reverberation and noise levels.}   
\end{abstract}

\begin{keywords}
sound source localization, \ac{RTF}, acoustic manifold, Gaussian process, \acf{ML}.
\end{keywords}

\section{Introduction}
\label{sec:intro}
Acoustic source localization is an essential component in various audio applications, such as: automated camera steering and teleconferencing systems~\cite{huang2000passive}, speaker separation~\cite{mandel2010model} and robot audition~\cite{nakadai2002real,valin2003robust,hornstein2006sound}. Thus, the localization problem has attracted a significant research attention, and a large variety of localization methods were proposed during the last decades. The main challenge facing the research community is how to perform robust localization in adverse conditions, namely, in the presence of background noise and reverberations, which are the main causes for performance degradation of localization algorithms. 

Broadly, traditional localization methods can be divided into three main categories: methods based on maximization of the \ac{SRP} of a beamformer output, high-resolution spectral estimation techniques, and dual-stage approaches relying on a \ac{TDOA} estimation. In the first category, the position is estimated directly from the measured signals after being filtered and summed together. Commonly, the \ac{ML} criterion is applied, which \BL{in the case of a single source}, culminates in inspecting the output power of a beamformer steered to different locations and in searching the points where it receives its maximum value~\cite{yao2002maximum}. The second category consists of high resolution methods, such as \ac{MUSIC}~\cite{schmidt1986multiple} and \ac{ESPRIT}~\cite{roy1989esprit} algorithms, that are based on the spectral analysis of the correlation matrix of the measured signals. In the third category, a dual stage approach is applied. In the first stage, the \acp{TDOA} of different pairs of microphones are estimated and collected. The different \ac{TDOA} readings correspond to single-sided hyperbolic \BLS{hyperplanes} (in 3D) representing possible positions. The intersection of these \BLS{hyperplanes} yields the estimated position. In this type of approaches the quality of the localization greatly depends on the quality of the \ac{TDOA} estimation in the first stage. The classical method for \ac{TDOA} estimation, which assumes a reverberant-free model, is the \acf{GCC} algorithm introduced in the landmark paper by Knapp and Carter~\cite{Knapp1976}. Many improvements of the \ac{GCC} method for the reverberant case were proposed, e.g. in~\cite{brandstein1997robust,stephenne1997new,rui2004time,dvorkind2005time,scheuing2008disambiguation}. \BLT{Among these methods for \ac{TDOA} estimation in reverberant conditions,} there are subspace methods based on adaptive eigenvalue decomposition~\cite{benesty2000adaptive} and generalized eigenvalue decomposition~\cite{doclo2003robust}. \BLT{ Of special importance is the \ac{SRP-PHAT} algorithm proposed in~\cite{dibiase2001robust}. This method is related to both the first and the third categories, since it combines in a single step the features of a steered-beamformer with those of the phase transform weighting of the \ac{GCC} algorithm.}            
 
Most of the traditional localization approaches are based on physical models and rely on certain assumptions regarding the propagation model and the statistics of the signals and the noise. However, in complex real-world scenarios, characterized by strong levels of noise and reverberation, a reliable model does not necessarily exist. Recently, there is a growing interest in learning-based localization approaches, which attempt to learn the characteristics of the acoustic environment directly from the data, in contrast to using a predefined physical model. Typically, these approaches assume that a training set of prerecorded measurements is given in advance. Supervised methods utilize microphone measurements of sources from known locations, while unsupervised approaches solely utilize the measurements, without knowing their exact source positions.  

\BLT{Learning-based approaches were proposed for both microphone array localization and binaural localization}. In the binaural hearing context, Deleforge and Horaud have proposed a probabilistic piecewise affine regression model that infers the localization-to-interaural \BL{data} mapping and its inverse~\cite{deleforge20122d}. They have extended this approach to the case of multiple sources using the variational \ac{EM} framework~\cite{deleforge2013variational,deleforge2015acoustic}.
In~\cite{may2011probabilistic}, another approach was presented based on a \ac{GMM} which was used to learn the azimuth-dependent distribution of the binaural feature space. In~\cite{Wu2016spatial}, a binaural localization method was proposed by assessing the mutual information between each of the spatial cues and the corresponding source location. In~\cite{xiong2015}, \ac{GCC}-based feature vectors were extracted and used for training a multilayer perceptron neural network that outputs the source \ac{DOA}. A method for \ac{DOA} estimation of multiple sources was presented in~\cite{xiong2016spatial}, using an \ac{EM} clustering approach. A localization method for a source located behind an obstacle that blocks the direct propagation path was presented in~\cite{kitic2014hearing}. The algorithm \BL{uses co-sparse data analysis based on the physical model of the wave propagation}. The model was extended in~\cite{bertin2016joint} to the case where the physical properties of the enclosure are not known in advance.

Talmon et al.~\cite{talmon2012parametrization} introduced a supervised method based on \emph{manifold learning}, aiming at recovering the fundamental controlling parameter of the \ac{AIR}, which coincides with the source position in a static environment. The method was applied to a single microphone system with a \ac{WGN} input~\cite{Talmon2011}.
In~\cite{laufer2013relative} we adopted the paradigm of~\cite{Talmon2011} and adapted it to a speech source, using a dual-microphone system with a \ac{PSD}-based feature vector. 
\BLT{Another approach for semi-supervised source localization with a single microphone pair, based on regularized optimization in a \ac{RKHS}, was recently presented in~\cite{laufer2016semi}.} 

\BLT{In this paper, we consider a setup consisting of multiple nodes, where each node comprises a pair of microphones. No additional assumptions, particularly on their specific (unknown) locations, are made. We believe that such an extension of the setup, which includes much more spatial information, is both practical and may lead to improved accuracy of localization tasks. In our recent work~\cite{laufer2016}, we reformulated the optimization problem presented in~\cite{laufer2016semi} using a Bayesian inference approach for the single node case. Following~\cite{sindhwani2005beyond,sindhwani2007semi}, we assumed that the function of interest, which attaches the position estimate to any measurement-based feature vector, follows a Gaussian process with a covariance function that is built based on a certain kernel function. This Bayesian framework enables us to naturally extend the single node setup to multiple nodes. Here as well, we focus on enclosures (such as car interiors, conference rooms, offices, etc.), which do not significantly change often, and thereby allow to establish a set of signal recordings in advance. In other words, we assume the availability of a training set consisting of a limited number of labelled measurements from multiple nodes, attached with corresponding source positions, and a larger amount of unlabelled measurements with unknown source locations. The unlabelled data is essential for identifying unique patterns and geometrical structures in the data, which are utilized for constructing data-driven models. The main idea is to define a Gaussian process with a new covariance function that encapsulates the connections between all available pairs of microphones, leveraging the information manifested in the acoustic samples acquired from different locations. In addition, this statistical framework allows for the rigorous estimation of the model parameters as an integral part of the optimization procedure, through an appropriate maximum likelihood (ML) criterion. Moreover, a recursive version is derived, where the new samples acquired during the test stage are utilized for updating the correlation of the process using an LMS-type approach.}

The paper is organized as follows. In Section~\ref{sec:problem}, we formulate the problem in a general noisy and reverberant environment. We discuss the existence of an acoustic manifold for each node and present the statistical model. A manifold-based Gaussian process is presented in Section~\ref{sec:statf}, and the relations between the nodes are defined. These definitions are unified by the \ac{MMGP} presented in Section~\ref{sec:fusion}, which combines together the information from all the nodes. Based on this model a Bayesian estimator is derived in Section~\ref{sec:interfernce}. We present a recursive adaptation mechanism, and describe how to estimate the model parameters using an \ac{ML} criterion. In Section~\ref{sec:results}, we demonstrate the algorithm performance by an extensive simulation study, and real-life recordings. Section~\ref{sec:conclusions} concludes the paper.

\section{Problem Formulation}
\label{sec:problem}
A single source is located in a reverberant enclosure at position $\mathbf{q}=[q_x,q_y,q_z]^T$. \BLT{Consider $M$ nodes consisting of pairs of microphones, distributed around the enclosure.} The source produces an unknown speech signal $s(t)$, which is measured by all the microphones. The signal received by the $i$th microphone of the $m$th pair, is given by: 
\begin{equation}
y^{m}_{i}(t)=a^{m}_{i}(t,\mathbf{q})*s(t)+u^m_i(t) \quad m=1,\ldots,M;\quad i=1,2 
\end{equation}
where $a^{m}_{i}(t,\mathbf{q})$ is the \acf{AIR} relating the source at position $\mathbf{q}$ and the $i$th microphone in the $m$th node, and $u^m_i(t)$ is an additive noise signal, which contaminates the corresponding measured signal. Linear convolution is denoted by $\ast$. 

Clearly, the information required for localization is \BLS{embedded} in the \ac{AIR} and is independent of the source signal. Thus, from each pair of measurements we extract a feature vector $\mathbf{h}^m$ that depends solely on the two \acp{AIR} of the corresponding node and is independent of the non-stationary source signal. More specifically, we use a feature vector based on \ac{RTF} estimates~\cite{Gannot2001} in a certain frequency band, which is commonly used in acoustic array processing~\cite{Gannot2001,markovich2009multichannel}. Please refer to~Appendix~\ref{sec:appA} for further details about the \ac{RTF} and its estimation. \BLT{The \acp{RTF} are typically represented in high dimension with a large number of coefficients to allow for the full description of the acoustic paths, which represent a complex reflection pattern. The observation that the \acp{RTF} are controlled by a small set of parameters, such as room dimensions, reverberation time, location of the source and the sensors etc., gives rise to the assumption that they are confined to a low dimensional manifold. In~\cite{laufer2015} and~\cite{laufer2016semi}, we have shown that the \acp{RTF} of a certain node have a distinct structure. Hence, they are not uniformly distributed in the entire space, but rather pertain to a manifold $\mathcal{M}_m$ of much lower dimensions.}

\BLS{We define the function $f^m_a: \mathcal{M}_m\rightarrow \mathbb{R} \quad a\in \{x,y,z\}$  which attaches the corresponding $x$,$y$ or $z$ coordinate of the source position $f^m_a(\mathbf{h}^m)$, to an \ac{RTF} sample $\mathbf{h}^m$ associated with the $m$th node. The three coordinates are assumed to be independent (proximity in one of the axes does not imply proximity in another axis), hence each coordinate is estimated individually. Since the same estimation is used for each coordinate, the axis notation is omitted henceforth. Let $p^m_l\equiv f^m(\mathbf{h}^m_l)$ denote the position evaluated by the function $f^m$ for the \ac{RTF} sample $\mathbf{h}^m_l$. In this notation, the superscript denotes association to a certain node, and the subscript denotes association to a certain position. Note that although the position of the source does not depend on the specific node, the notation $p^m_l$ is used to express that the mapping is obtained from the measurement of the $m$th node.}

The $m$th \ac{RTF} represents the reflection pattern originating from the source and received by the $m$th node. Assuming that the different nodes are scattered over the room area, they experience a distinct reflection pattern which differs from that experienced by other nodes. Each \ac{RTF} $\mathbf{h}^m$ represents a different view point on the same acoustic event of a source speaking at some location in the enclosure. A particular node may have an accurate view of certain regions in the room and yet lacking on others. For example, closer distances are better viewed, while remote positions are not well distinguished. The view point of each node is reflected by the manifold $\mathcal{M}_m$ whose structure represents the relations between different \acp{RTF}, as they are inspected by that node. Combining the information from the different nodes may therefore increase the spatial separation and improve the ability to accurately locate the source. The central issue is then how to fuse the information provided by each of the $M$ nodes to achieve this goal.   

\BLS{Let $\mathbf{h}=\left[[\mathbf{h}^1]^T, \ldots ,[\mathbf{h}^M]^T\right]^T$ denote the \ac{aRTF}, which is a concatenation of the \ac{RTF} vectors from every pair of microphones. We define the scalar function $f: \cup_{m=1}^M\mathcal{M}_m\rightarrow \mathbb{R}$  which attaches an \ac{aRTF} sample $\mathbf{h}_l$ with the corresponding $x$,$y$ or $z$ coordinate of the source position $p_l\equiv f(\mathbf{h}_l)$. In the first step, we discuss each node and its mapping function $f^m$, and then we combine the different views in the definition of the function $f$. The estimation of the function is semi-supervised and is based on a set of \ac{aRTF} samples associated with various source positions, measured in advanced. However, the microphone positions may be unknown since they are not required for the estimation. The training set consists of two subsets: a small subset of samples with ``labels", i.e. with known source positions, and a large subset of \ac{aRTF} samples without labels, i.e., with unknown source locations. Since all the samples in this set correspond to measurements from the same enclosure, we assume that they are confined to the same manifold. }  

The relation between the $m$th \ac{RTF} sample and the associated position is dictated by the specific acoustic environment to be inspected, i.e. surfaces materials, room dimensions and microphone locations. In a fixed acoustic environment, the function $f^m$ that relates $\mathbf{h}^m_l$ to its position $p_l^m$ \BL{(which is a scalar since it represents the $x$,$y$ or $z$ coordinate of the position)}, is deterministic, in the sense that a certain reflection pattern \BL{expressed by} the $m$th \ac{RTF} is exclusively  associated with a certain position. However, even when all the environmental parameters are fixed and known, there is no simple model that links a given \ac{RTF} sample to its position. Hence, we use the statistical model presented in~\cite{laufer2016}. An \ac{RTF} $\mathbf{h}^m_l$ is assumed to be sampled from the manifold $\mathcal{M}_m$. \BLT{The \ac{RTF} sample $\mathbf{h}^m_l$ is related by the function $f^m$ to the corresponding position $p^m_l$. We assume that $p^m_l$ is a realization of a stochastic process. \BLS{The physical positions of the source are measured for the labelled training samples, which serve as their corresponding labels}. This yields a noisy version $\bar{p}_l$ of the actual position, due to imperfections in the measurements.}

\BLT{Before describing the algorithm, we reiterate the definition of the entire set of all the measurements, which consists of a training set used for advanced learning, and a test set for which a position estimation is required. As mentioned before, the training set consists of two subsets. The first subset consists of $n_L$ labelled samples, denoted by $H_L=\{\mathbf{h}_i\}_{i=1}^{n_L}$, with associated \BL{noisy} labels $P_L=\{\bar{p}_i\}_{i=1}^{n_L}$. \BLS{Note that though all three coordinates of the position are measured for each labelled sample, $P_L$ is defined as a collection of scalars (associated with a certain coordinate) rather than vectors, since the same derivation applies separately to each coordinate.} The second subset consists of $n_U$ unlabelled samples, denoted by $H_U=\{\mathbf{h}_i\}_{i=n_L+1}^{n_D}$, where $n_D=n_L+n_U$. The entire training set consists of $n_D$ \ac{aRTF} samples and is denoted by $H_D=H_L\cup H_U$. In the test stage, we receive a new set $H_T=\{\mathbf{h}_i\}_{i=n_D+1}^n$ of $n_T$ new \ac{aRTF} samples from unknown locations, where $n=n_D+n_T$. The entire set, including both the training and the test samples, is denoted by $H=H_D \cup H_T$.}

\section{Manifold-Based Gaussian Process}
\label{sec:statf} 
\BLS{We first present the statistical model for each node individually, and then discuss the relations between different nodes. Finally, we define the function $f$ that combines the data from all the nodes in a way that respects both the interior relations within each node and the inter-relations between the different nodes.}

\BL{We assume that the position $p^m$, which is the output of the function associated with the $m$th node, follows a Gaussian process, i.e. the set of all possible positions mapped from the samples of the $m$th pair, are jointly distributed Gaussian variables.} The Gaussian process is a convenient choice since it is entirely defined by its second order statistics, and is widely used for regression problems~\cite{rasmussen2006gaussian}. \BL{We use a zero-mean Gaussian process} for simplicity. However, all the results apply also to any general mean function with only minor changes. The covariance function is a pairwise affinity measure between two \ac{RTF} samples. We suggest to use a manifold-based covariance function in which the relation between two \acp{RTF} is not only a function of the current samples, but also uses information from the entire available set of \ac{RTF} samples:
\begin{multline}
\textrm{cov}(p^m_r,p^m_l)\equiv \sum_{i=1}^{n_D}k_m(\mathbf{h}^m_r,\mathbf{h}^m_i)k_m(\mathbf{h}^m_l,\mathbf{h}^m_i)\\ \numberthis \label{eq:cov_man}
=2k_m(\mathbf{h}^m_r,\mathbf{h}^m_l)+\sum_{i=1\atop i\ne l,r}^{n_D}k_m(\mathbf{h}^m_r,\mathbf{h}^m_i)k_m(\mathbf{h}^m_l,\mathbf{h}^m_i) 
\end{multline}
\BL{where $l$ and $r$ represent ascription to certain positions}, and $k_m$ is a standard pairwise function $k_m: \mathcal{M}_m\times \mathcal{M}_m\longrightarrow \mathbb{R}$, often termed ``kernel function''. \BL{The equality in~\eqref{eq:cov_man} holds for kernels that satisfy: $k_m(\mathbf{h}^m_i,\mathbf{h}^m_j)=1$ for $i=j$}. A common choice is to use a Gaussian kernel, with a scaling factor $\varepsilon_m$: 
\begin{equation}
k_m(\mathbf{h}^m_i,\mathbf{h}^m_j)=\exp\left\{-\frac{\Vert\mathbf{h}^m_i-\mathbf{h}^m_j\Vert^2} {\varepsilon_m} \right\}.
\label{eq:gauss}
\end{equation} 
The definition of the covariance in~\eqref{eq:cov_man}, induces a new type of manifold-based kernel $\tilde{k}_m$:
\begin{equation}
\tilde{k}_{m}(\mathbf{h}^m_r,\mathbf{h}^m_l) \equiv \textrm{cov}(p^m_r,p^m_l) 
\end{equation} 
In~\cite{laufer2016} we adopted the manifold-based kernel proposed by Sindhwani et al.~\cite{sindhwani2007semi}. Here, we propose another type of kernel, which is more convenient for estimating the model hyperparameters and for deriving a recursive adaptation mechanism. A similar kernel was used to define a graph-based diffusion filter in~\cite{kushnir2012anisotropic} and was applied in a patch-based de-noising algorithm in~\cite{haddad2014texture}. Note that the new kernel $\tilde{k}_{m}$ consists of the standard kernel $k_m$ and a second term that represents the mutual correlation between the two \ac{RTF} samples when compared to all other existing samples as viewed by the $m$th node. When $\mathbf{h}^m_i$ and $\mathbf{h}^m_j$ are mutually close to the same subset of \ac{RTF} samples, it indicates that they are closely related with respect to the manifold, and the value of $\tilde{k}(\mathbf{h}^m_r, \mathbf{h}^m_l)$ increases respectively. In general, we can state that the second term in~\eqref{eq:cov_man} compares between the embeddings of $\mathbf{h}^m_r$ and $\mathbf{h}^m_l$ in the manifold $\mathcal{M}_m$ and updates the correlation between the two accordingly. Since the manifold-based kernel $\tilde{k}_m$ takes into consideration the relations to other samples from the manifold it may be preferable over the standard kernel $k_m$ in~\eqref{eq:gauss}. 

\begin{figure}[ht!]
\centering
\includegraphics[width=0.45\textwidth,height=0.23\textheight]{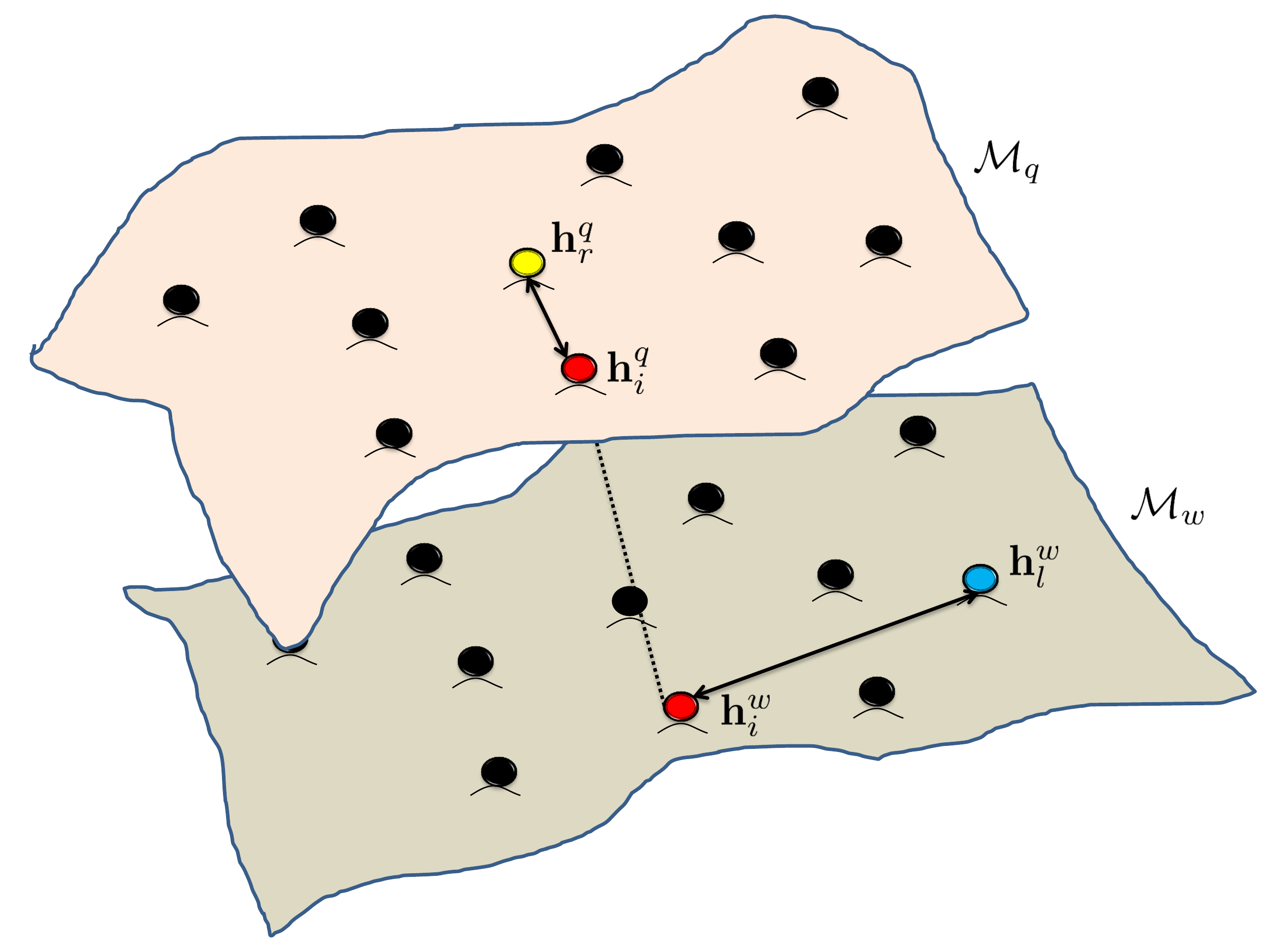} 
\caption{An illustration of the covarience computation for \ac{RTF} samples of different nodes $q$ and $w$} \label{fig:exter}
\end{figure}

\BLS{Similarly, we define the relation between the functions of different nodes $q$ and $w$, evaluated for two \ac{RTF} samples associated with different source positions. Namely, we define the relation between $p^q_r$ and  $p^w_l$ for $1\leq l,r\leq N_D$.} We assume that $p^q_r$ and  $p^w_l$ are jointly Gaussians and that their covariance is defined by:
\BL{\begin{align*}
\textrm{cov}(p^q_r,p^w_l)&\equiv\tilde{k}_{qw}(\mathbf{h}^q_r,\mathbf{h}^w_l)\\
&=\sum_{i=1}^{n_D}k_q(\mathbf{h}^q_r,\mathbf{h}^q_i)k_w(\mathbf{h}^w_l,\mathbf{h}^w_i). \numberthis
\label{eq:cross}
\end{align*}}
It is important to note that when examining the relation between functions evaluated for different nodes, we cannot directly compute the distance between the corresponding \ac{RTF} samples since they represent different views. In~\eqref{eq:cross}, we propose to choose another sample $\mathbf{h}_i$ associated with a certain source position, and \BL{compare the inter-relations in the $q$th manifold between $\mathbf{h}^q_r$ and $\mathbf{h}^q_i$, and the inter-relations in the $w$th manifold between $\mathbf{h}^w_l$  and $\mathbf{h}^w_i$, as illustrated in Fig.~\ref{fig:exter}}. 

\BLT{\section{Multi-Node Data Fusion}}
\label{sec:fusion} 
So far, we have presented the statistical model and defined a Gaussian process $p^m$ for each node. In addition, we have defined the covariance of each individual process of a particular node~\eqref{eq:cov_man} and the cross-covariance between two processes of two different nodes~\eqref{eq:cross}. Our goal is to unify these definitions under one statistical umbrella which combines the information provided by the different pairs and establishes a foundation for deriving a Bayesian estimator for the source position. 

\BLT{\subsection{Multiple-Manifold Gaussian Process}}  
To fuse the different perspectives presented by the different nodes, we define the \BL{\acf{MMGP}} $p$ as the mean of the Gaussian processes of all the nodes:
\begin{equation}
p=\frac{1}{M}(p^1+p^2+\ldots+p^M).
\end{equation}    
Due to the assumption that the processes are jointly Gaussian, the \BLT{process $p$} is also Gaussian with zero-mean and covariance function given by:
\begin{align*}
\textrm{cov}(p_r,p_l)&=\frac{1}{M^2}\textrm{cov}\left(\sum_{q=1}^M p_r^q,\sum_{w=1}^M p_l^w\right)\\ 
&=\frac{1}{M^2}\sum_{q,w=1}^M \textrm{cov}(p^q_r,p^w_l). \numberthis  
\label{eq:MMGP_cov}
\end{align*}
Using the definitions of~\eqref{eq:cov_man} and~\eqref{eq:cross} we obtain the covariance \BLT{for} $p_r$ and $p_l$:
\begin{align*}
\textrm{cov}(p_r,p_l)&\equiv\tilde{k}(\mathbf{h}_r,\mathbf{h}_l)\\
&=\frac{1}{M^2}\sum_{i=1}^{n_D}\sum_{q,w=1}^M k_q(\mathbf{h}^q_r,\mathbf{h}^q_i)k_w(\mathbf{h}^w_l,\mathbf{h}^w_i). \numberthis
\label{eq:sum_cov}
\end{align*} 
Here, the covariance, \BLT{evaluated for} two samples from the process $p$, is determined using all $M^2$ relations between the different nodes and by averaging over all the samples in $H_D$. \BLS{Through the lens of kernel-based learning,} $\tilde{k}(\mathbf{h}_r,\mathbf{h}_l)$ can be considered as a \emph{composition of kernels}, which, in addition to connections acquired in each node separately, incorporates the extra spatial information manifested in the mutual relationship between \acp{aRTF} from the different nodes. This formulation represents a robust measurement of correlation by utilizing multiple view-points of the same acoustic scene, aiming to improve the localization capabilities.

The resulting Gaussian process is zero-mean with covariance function $\tilde{k}$:
\begin{equation}
p \thicksim \mathcal{GP}(0,\tilde{k}).
\label{eq:MGP}
\end{equation} 
Accordingly, the random vector $\mathbf{p}_H=[p_1,\ldots,p_n]^T$, which consists of $n$ samples from the process $p$, has a multivariate Gaussian distribution, i.e.,
\begin{equation}
\mathbf{p}_H \sim \mathcal{N}(\mathbf{0}_{n},\tilde{\mathbf{\Sigma}}_{H})
\label{eq:prior}
\end{equation} 
where $\mathbf{0}_{n}$ is an $n\times 1$ vector of all zeros and $\tilde{\mathbf{\Sigma}}_{H}$ is the covariance matrix with elements $\tilde{k}(\mathbf{h}_i,\mathbf{h}_j),\;\mathbf{h}_i,\mathbf{h}_j\in H$. Note that the covariance matrix $\tilde{\mathbf{\Sigma}}_{H}$ can be expressed in terms of the covariance matrices of all the individual nodes $\mathbf{K}^m_H$, defined by the standard kernel $(\mathbf{K}^m_H)_{ij}=k_m(\mathbf{h}^m_i,\mathbf{h}^m_j)$ of~\eqref{eq:gauss}:
\begin{equation}
\tilde{\mathbf{\Sigma}}_{H}=\frac{1}{M^2}\sum_{q,w=1}^M\mathbf{K}^q_{H}\mathbf{K}^w_{H}.
\label{eq:pairs}
\end{equation}  
In this representation, the covariance matrix for any finite set of samples from the process is computed by a sum of all pairwise multiplications between the covariance matrices of each of the nodes.

\subsection{Alternating Diffusion Interpretation}
\label{sec:alternate}
Before we proceed to the derivation of the estimation procedure which is based on these definitions, \BL{we present an alternative interpretation} using a geometrical perspective from the field of diffusion maps~\cite{Coifman2006}. Specifically, we provide an interpretation for the definitions of the covariance functions in~\eqref{eq:cov_man}, \eqref{eq:cross} and \eqref{eq:sum_cov}. As discussed above, every node represents a different view point, which is realized by the structure of the associated manifold $\mathcal{M}_m$. We can create a discrete representation of the $m$th manifold by a graph $G^m$ in which the vertices represent the \ac{RTF} samples of the $m$th node and the weights connecting between them are stored in the matrix $\mathbf{K}^m_H$. This way, we obtain $M$ graphs \BLT{with matching vertices that are associated with the same positions}, but with different weighted edges determined by the distances between the samples within each separate node. In~\cite{lederman2015learning}, the authors defined an alternating diffusion operator, which constitutes a combined graph $G^{qw}$ where the weight matrix is given by $\mathbf{K}_H^{qw}\equiv\mathbf{K}^q_H\mathbf{K}^w_H$. They have shown that the Markov process defined on the resulting graph extracts the underlying source of variability common to the \BLT{two graphs $q$ and $w$ (related to the microphone nodes $q$ and $w$)}. 

In our case, an \ac{RTF} is closely related to its associated position, however it may be influenced by other factors as well, such as estimation errors and noise. We assume that the interferences introduced by a particular node differ from the ones introduced by the other nodes. When measuring the correlation between two nodes, we would like to emphasize the common source of variability, namely the source position, and to suppress artifacts and interferences which are node-specific effects. By multiplying the kernels of each two nodes as indicated in~\eqref{eq:pairs}, we average out incoherent node-specific variables and remain only with the common variable which is the position of the source. This perspective provides a justification to the averaging over different nodes as well as over different samples, constituting a robust measure of correlation between samples in terms of the physical proximity between the corresponding source positions. 
   
\section{Bayesian Inference with Multiple-Manifold Gaussian Process}
\label{sec:interfernce}
In the previous section we presented the \ac{MMGP} $p$ that relates the \ac{aRTF} to the corresponding source position. We have shown that the covariance of the process depends on both the internal relations within the same manifold (same node) and the pairwise connections between different manifolds (different nodes). Note that the covariance function of the process~\eqref{eq:sum_cov} is based only on the \ac{RTF} samples in $H_D$, and does not take into account the labellings. The information implied by the labelled samples $H_L$ and their associated labels $P_L$ is used to update our prior belief about the behaviour of the process $p$ and to derive its posterior distribution. The pairs $\{\mathbf{h}_i,\bar{p}_i\}_{i=1}^{n_L}$ serve as anchor points utilized for interpolating a realization of the process $p$, while the Gaussian process assumption in~\eqref{eq:MGP} is designed to ensure the smoothness of the solution.

\subsection{Localization with Multiple-Manifold Gaussian Process}
Following the statistical model stated in Section~\ref{sec:problem}, we assume that the measured positions $P_L=\{\bar{p}_i\}_{i=1}^{n_L}$ of the labelled set arise from a noisy observation model, given by:
\begin{equation}
\bar{p}_i=p_i+\eta_i; \mbox{ } i=1,\ldots,n_L
\label{eq:observ_local}
\end{equation}
where $\eta_i\sim\mathcal{N}(0,\sigma^2) \mbox{ } i=1,\ldots,n_L$ are i.i.d. Gaussian noises, independent of $p_i$. The noise in~\eqref{eq:observ_local} reflects uncertainties due to imperfect measurements of the source positions while acquiring the labelled set. Note that since the Gaussian variables $p_i$ and $\eta_i$ are independent, they are jointly Gaussian. Consequently, $p_i$ and  $\bar{p}_i$ are also jointly Gaussian. We define \BL{the likelihood of the process $p$} based on the probability of the labelled examples: 
\begin{equation}
\textrm{Pr}(P_L|p,H_L)=\frac{1}{\sqrt{2\pi\sigma^2}}\exp\left\{-\frac{1}{2\sigma^2}\sum_{i=1}^{n_L}(\bar{p}_i-p_i)^2\right\}.
\label{eq:ll}
\end{equation}   

To perform localization, we are interested in estimating the position of a new test sample $\mathbf{h}_{t} \in H_T$ of an unknown source from an unknown location. The estimation is based on the posterior probability $\textrm{Pr}(p_t\equiv f(\mathbf{h}_t)| P_L, H_L)$. According to~\eqref{eq:prior} and~\eqref{eq:ll}, the function value at the test point $p_t$ and the concatenation of all labelled training positions $\mathbf{p}_L=\textrm{vec}\{P_L\}\equiv[\bar{p}_1,\ldots,\bar{p}_{n_L}]^T$ are jointly Gaussian, with:
\begin{equation}
\begin{bmatrix}
      \mathbf{p}_L \\
      p_t
\end{bmatrix} \bigg| H_L
\sim
\mathcal{N}\left(
\mathbf{0}_{l+1}
,
\begin{bmatrix}
      \tilde{\mathbf{\Sigma}}_{L}+\sigma^2\mathbf{I}_{n_L} & \tilde{\mathbf{\Sigma}}_{Lt} \\
     \tilde{\mathbf{\Sigma}}_{Lt}^T & \tilde{\Sigma}_{t}
\end{bmatrix}
\right)
\label{eq:normd}
\end{equation}
where $\tilde{\mathbf{\Sigma}}_{L}$ is an $n_L\times n_L$ covariance matrix defined over the function values at the labelled samples $H_L$, $\tilde{\bold{\Sigma}}_{Lt}$ is an $n_L\times 1$ covariance vector between the function values at $H_L$ and $f(\mathbf{h}_t)$, $\tilde{\Sigma}_{t}$ is the variance of $p_t$, and $\mathbf{I}_{n_L}$ is the $n_L\times n_L$ identity matrix. 
This implies that the conditional distribution $\textrm{Pr}(p_t|P_L,H_L)$ is a multivariate Gaussian with $\mu_\textrm{cond}$ mean and $\sigma^2_\textrm{cond}$ variance given by:
\begin{align*}
\mu_\textrm{cond}&=\tilde{\mathbf{\Sigma}}_{Lt}^T\left(\tilde{\mathbf{\Sigma}}_{L}+\sigma^2\mathbf{I}_{n_L}\right)^{-1}\mathbf{p}_L \label{eq:cond}\\
\sigma^2_\textrm{cond}&=\tilde{\Sigma}_{t}-\tilde{\mathbf{\Sigma}}_{Lt}^T\left(\tilde{\mathbf{\Sigma}}_{L}+\sigma^2\mathbf{I}_{n_L}\right)^{-1}\tilde{\mathbf{\Sigma}}_{Lt}. \numberthis
\end{align*}
Hence, the \ac{MAP} estimator of $p_t$, which coincides with the \ac{MMSE} estimator in the Gaussian case, is given by:
\begin{equation}
\widehat{p}_t=\mu_\textrm{cond}=\tilde{\mathbf{\Sigma}}_{Lt}^T\tilde{\mathbf{p}}_L
\label{eq:est}
\end{equation}
\BL{where $\tilde{\mathbf{p}}_L\equiv\mathbf{\Gamma}_{L}\mathbf{p}_L$ is a vector of weights which are independent of the current test sample, and $\mathbf{\Gamma}_{L}=\left(\tilde{\mathbf{\Sigma}}_{L}+\sigma^2\mathbf{I}_{n_L}\right)^{-1}$}. Note that the estimator in~\eqref{eq:est} is obtained as a linear combination of the kernel $\tilde{k}$ evaluated for the test sample $\mathbf{h}_t$ and each of the labelled samples $H_L$, weighted by the entries of $\tilde{\mathbf{p}}_L$. \BL{Note that the posterior is defined only with respect to the labelled samples, hence the covariance terms are calculated based solely on the labelled samples $H_L$, without taking into account the samples in the set $H$ as was defined in general in the previous section}. Although the unlabelled samples do not appear explicitly in~\eqref{eq:est}, they take role in the computation of the correlation terms \BLS{as implied by~\eqref{eq:sum_cov}}. In fact, the unlabelled samples are essential both for obtaining a more accurate computation of the weights $\tilde{\mathbf{p}}_L$, and for better quantifying the relations between the current test sample and each of the labelled samples. The variance of the estimator is given by $\sigma^2_\textrm{cond}$ in~\eqref{eq:cond}. It can be seen that the posterior variance $\sigma^2_\textrm{cond}$ is smaller than the prior variance $\tilde{\Sigma}_{t}$, indicating that the labelled examples reduce the uncertainty in the behaviour of the Gaussian process. The variance of the estimator is smaller for test samples which are close to a large number of labelled samples, increasing the second term in~\eqref{eq:cond}, and therefore decreasing the overall variance. The estimation is more reliable in regions where the labelled samples are dense, and becomes more uncertain in sparse regions.   

\subsection{Recursive Algorithm}
\label{sec:recurs}
\BL{In this section, we develop a recursive version for the estimator in~\eqref{eq:est}. The Gaussian process is adapted by the information provided by new (streaming) \ac{RTF} samples, in the test stage. Any new \ac{RTF} sample $\mathbf{h}_t$ can be considered as an additional unlabelled sample, hence can be used to update the covariances in~\eqref{eq:cov_man} and \eqref{eq:cross}. Taking the new sample into consideration, the covariance in~\eqref{eq:sum_cov} for two labelled samples $1\leq l,r\leq n_L$, is updated by:
\begin{align*}
\tilde{k}^*(\mathbf{h}_r,\mathbf{h}_l)
&=\frac{1}{M^2}\sum_{i=1}^{n_D}\sum_{q,w=1}^M k_q(\mathbf{h}^q_r,\mathbf{h}^q_i)k_w(\mathbf{h}^w_l,\mathbf{h}^w_i)\\
&+\frac{1}{M^2}\sum_{q,w=1}^M k_q(\mathbf{h}^q_r,\mathbf{h}^q_t)k_w(\mathbf{h}^w_l,\mathbf{h}^w_t)\\
&=\tilde{k}(\mathbf{h}_r,\mathbf{h}_l)\\
&+\frac{1}{M^2}\left(\sum_{q=1}^Mk_q(\mathbf{h}^q_r,\mathbf{h}^q_t)\right)\left(\sum_{w=1}^Mk_w(\mathbf{h}^w_l,\mathbf{h}^w_t)\right) \numberthis
\label{eq:upd_cov}
\end{align*}
where $^*$ stands for an updated term. \BLS{Thus, the updated covariance defined over the labelled samples is given by a rank-1 update:}
\begin{equation}
\tilde{\mathbf{\Sigma}}^*_{L}=\tilde{\mathbf{\Sigma}}_{L}+\frac{1}{M^2}\mathbf{k}_{Lt}\mathbf{k}^T_{Lt}
\label{eq:upd_sigL}
\end{equation}
where $\mathbf{k}_{Lt}=\left[\sum_{q=1}^Mk_q(\mathbf{h}^q_1,\mathbf{h}^q_t),\ldots,\sum_{q=1}^Mk_q(\mathbf{h}^q_{n_L},\mathbf{h}^q_t)\right]^T$. Note that the updated correlation in~\eqref{eq:upd_cov}, when measured between the new test sample $\mathbf{h}_t$ and a labelled sample $\mathbf{h}_l$, is given by $\tilde{k}^*(\mathbf{h}_t,\mathbf{h}_l)=\tilde{k}(\mathbf{h}_t,\mathbf{h}_l)+\frac{1}{M}\sum_{q=1}^Mk_q(\mathbf{h}^q_l,\mathbf{h}^q_t)$ for kernels that satisfy $k_m(\mathbf{h}^m_i,\mathbf{h}^m_j)=1$ for $i=j$. Hence, the updated covariance vector between the new test sample and each of the labelled samples is given by:
\begin{equation}
\mathbf{\tilde{\Sigma}}^*_{Lt}=\tilde{\mathbf{\Sigma}}_{Lt}+\frac{1}{M}\mathbf{k}_{Lt}
\end{equation}  
\BLS{Using the Woodbury matrix identity~\cite{woodbury1950inverting}} and~\eqref{eq:upd_sigL}, we obtain the adaptation rule for $\mathbf{\Gamma}_{L}$:
\begin{align*}
\mathbf{\Gamma}^*_{L}&=\left(\mathbf{\Gamma}^{-1}_{L}+\frac{1}{M^2}\mathbf{k}_{Lt}\mathbf{k}^T_{Lt}\right)^{-1}\\
&=\mathbf{\Gamma}_{L}-\frac{\mathbf{\Gamma}_{L}\mathbf{k}_{Lt}\mathbf{k}^T_{Lt}\mathbf{\Gamma}_{L}}{M^2+\mathbf{k}^T_{Lt}\mathbf{\Gamma}_{L}\mathbf{k}_{Lt}} \numberthis
\label{eq:recurs_cov}
\end{align*}  
\BLS{where the new sample is utilized to form a more accurate measure of the correlation between the labelled samples.} }Hence, the updated weights are $\tilde{\mathbf{p}}^*_L=\mathbf{\Gamma}^*_{L}\mathbf{p}_L$, and the estimated position is given by:
\begin{equation}
\hat{p}_t=\mathbf{\tilde{\Sigma}}^{*T}_{Lt}\tilde{\mathbf{p}}^*_L.
\label{eq:recurs_est}
\end{equation}

\subsection{Learning the Hyperparameters}
\label{sec:parameters}
The zero-mean Gaussian process model is fully specified by its covariance function. Thus, the predictions obtained by this model depend on the chosen covariance function. \BLS{In practice, we use a parametric family of functions, i.e. a Gaussian kernel as in~\eqref{eq:gauss} with a scaling-parameter $\varepsilon_m$. The values of the parameters $\{\varepsilon_m\}_{m=1}^M$ can be inferred from the data by optimizing the likelihood function of the labelled samples. From the distribution defined in~\eqref{eq:normd}, the log-likelihood function of the labelled samples get the form of a multivariate Gaussian distribution, given by:}            
\begin{multline}
L=\ln \textrm{Pr}(\mathbf{p}_L|H_L;\Theta)
=-\frac{1}{2}\mathbf{p}_L^T\left(\tilde{\mathbf{\Sigma}}_{L}+\sigma^2\mathbf{I}_{n_L}\right)^{-1}\mathbf{p}_L\\
-\frac{1}{2}\ln\left|\tilde{\mathbf{\Sigma}}_{L}+\sigma^2\mathbf{I}_{n_L}\right|-\frac{{n_L}}{2}\ln(2\pi)
\label{eq:ll_f}
\end{multline}
\BL{where the first term measures how well the parameters fit the given labelled samples and the second term reflects the model complexity which is evaluated through the determinant of the covariance matrix.}
The optimization requires the computation of the gradients of the log-likelihood function with respect to each of the parameters. The partial derivative with respect to $\varepsilon_m$ can be generally expressed by \BL{(see~\cite{rasmussen2006gaussian} Chapter 5)}:
\begin{align*}
\frac{\partial L}{\partial \varepsilon_m}&=-\frac{1}{2}\textrm{trace}\left\{\mathbf{\Gamma}_{L}\frac{\partial \tilde{\mathbf{\Sigma}}_{L}}{\partial \varepsilon_m}\right\}+\frac{1}{2}\mathbf{p}_L^T\mathbf{\Gamma}_{L}\frac{\partial \tilde{\mathbf{\Sigma}}_{L}}{\partial \varepsilon_m}\mathbf{\Gamma}_{L}\mathbf{p}_L\\
&=\frac{1}{2}\textrm{trace}\left\{\left[(\mathbf{\Gamma}_{L}\mathbf{p}_l)(\mathbf{\Gamma}_{L}\mathbf{p}_l)^T-\mathbf{\Gamma}_{L}\right]\frac{\partial \tilde{\mathbf{\Sigma}}_{L}}{\partial \varepsilon_m}\right\} \numberthis
\label{eq:deriv}
\end{align*}
where the partial derivative of $\mathbf{\tilde{\Sigma}}_{L}$ in~\eqref{eq:deriv} with respect to each $\varepsilon_m$, is given by: 
\begin{align*}
M^2\frac{\partial \tilde{\mathbf{\Sigma}}_{L}}{\partial \varepsilon_m}&= \frac{\partial \left(\sum_{q,w=1}^M\mathbf{K}^q_{L} \mathbf{K}^w_{L}\right)}{\partial \varepsilon_m}\\
&=\frac{\partial \mathbf{K}^m_{L}}{\partial \varepsilon_m}\left(\sum_{q=1}^M\mathbf{K}^q_{L}\right)
+\left(\sum_{q=1}^M\mathbf{K}^q_{L}\right)\frac{\partial \mathbf{K}^m_{L}}{\partial \varepsilon_m} \numberthis
\label{eq:deriv_mat}
\end{align*} 
where $\frac{\partial \mathbf{K}^m_{L}}{\partial \varepsilon_w}$ is an $n_L\times n_L$ matrix with $(i,j)$th entry given by $\frac{\Vert\mathbf{h}_i-\mathbf{h}_j\Vert^2} {\varepsilon_m^2}\exp\left\{-\frac{\Vert\mathbf{h}_i-\mathbf{h}_j\Vert^2} {\varepsilon_m} \right\}$ .

\begin{figure*}[ht!]
\centering
\includegraphics[width=0.75\textwidth,height=0.4\textheight]{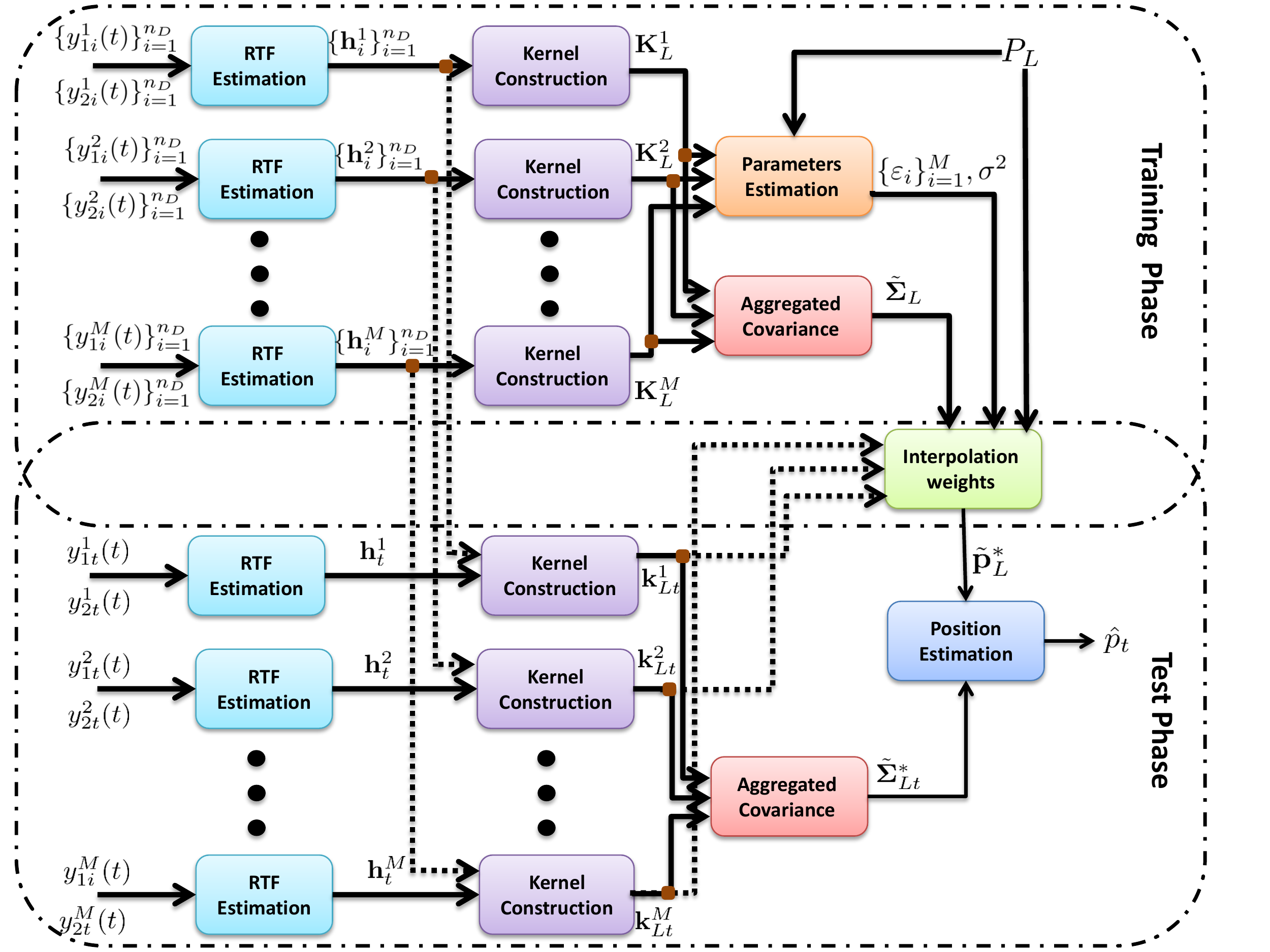} 
\caption{A flow diagram of the proposed algorithm} \label{fig:flow}
\end{figure*}

Similarly, we can also estimate the optimal value for the variance $\sigma^2$ of the observation noise. The partial derivative with respect to $\sigma^2$ has similar form to~\eqref{eq:deriv}:
\begin{equation}
\frac{\partial L}{\partial \sigma^2}=\frac{1}{2}\textrm{trace}\left\{(\mathbf{\Gamma}_{L}\mathbf{p}_l)(\mathbf{\Gamma}_{L}\mathbf{p}_l)^T-\mathbf{\Gamma}_{L}\right\}.
\label{eq:deriv_sig}
\end{equation}  
\BLS{Based on~\eqref{eq:deriv},~\eqref{eq:deriv_mat} and~\eqref{eq:deriv_sig}, Eq.~\eqref{eq:ll_f} can be optimized using an efficient gradient-based optimization algorithm.} \BLT{It should be noted that the parameter values are optimized through the likelihood of the labelled set, hence, optimality for the test samples cannot be guaranteed. This optimization can serve as an initialization for the parameter values, which may then be fine-tuned by other prevailing methods such as cross-validation.}
A flow diagram of the entire algorithm is illustrated in Fig.~\ref{fig:flow}. 

\section{Experimental Results}
\label{sec:results}
In this section, we demonstrate the performance of the proposed method for localization of a single source in noisy and reverberant conditions. We focus on 2-dimensional localization in both the $x$ and the $y$ coordinates. However, the algorithm can be applied to full 3-dimensional localization as well. The performance is evaluated using both simulated data and real-life recordings. The simulation is used to give a wide comparison of the effect of different noise and reverberation levels. However, the examination of real recordings is of great importance, since the simulation may not faithfully represent the physical phenomenons encountered in real-life scenarios. 

\subsection{Simulation Results}
\begin{figure}[ht!]
\centering
\includegraphics[trim=1cm 8cm 1cm 9cm, width=0.5\textwidth,height=0.25\textheight]{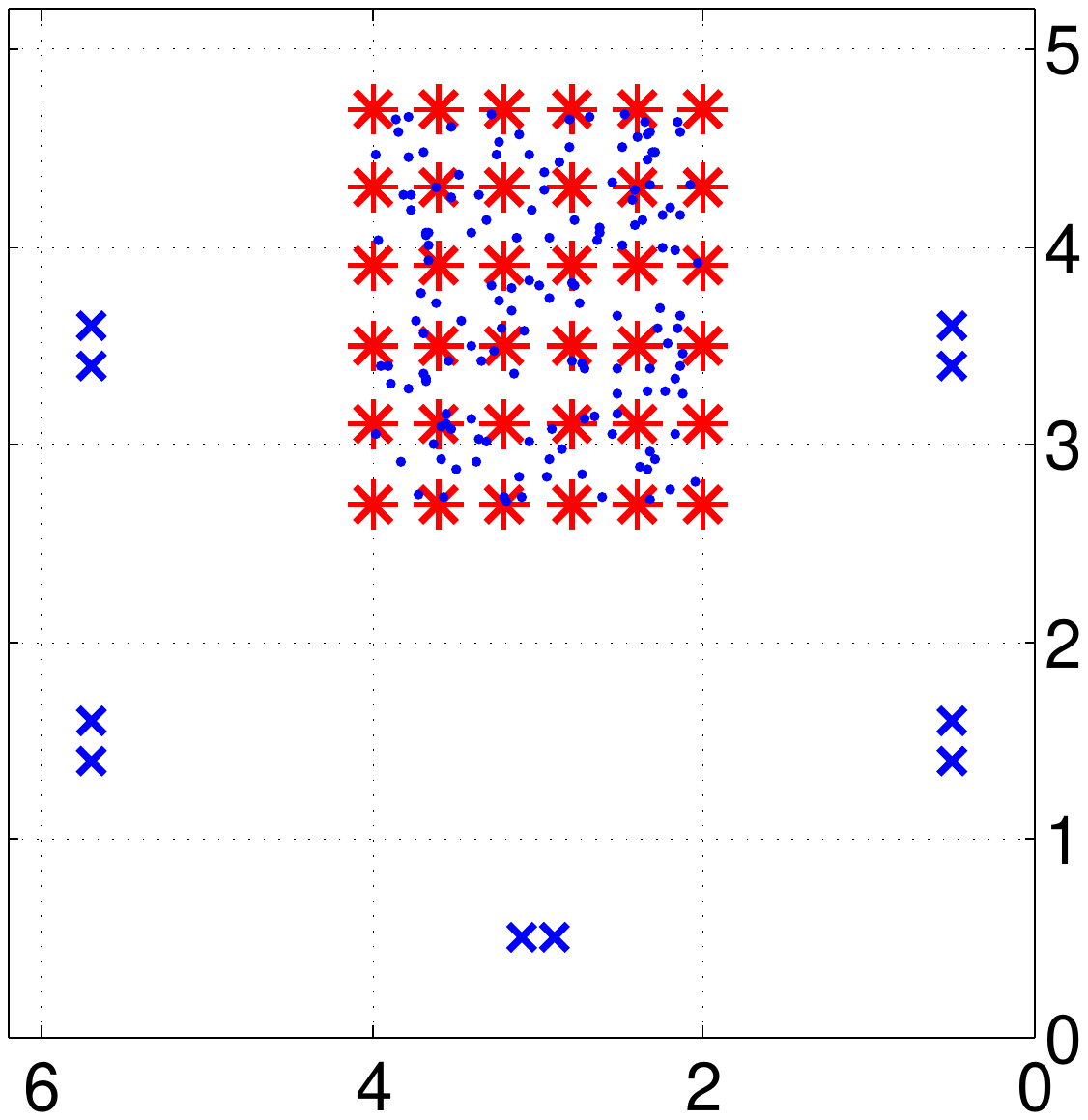}
\centering
\caption{The room setup. The blue x-marks denote the microphones, the red asterisks denote the labelled samples and the blue dots denote the unlabelled samples.} 
\label{fig:sim_setup}
\end{figure}

We simulated a $6\times 6.2\times 4$m room with different reverberation levels, using an efficient implementation~\cite{Habets2006} of the image method~\cite{Image79}. Six pairs of microphones are located around the room. The source  positions are confined to a $2\times2$m squared region, at $0.5$~m distance from one of the room walls. The training set consists of $n_L=36$ labelled samples creating a grid with a resolution of  $40$cm. In addition, there are $n_U=150$ unlabelled measurements from unknown locations in the same region. The room setup and the positions of the training set are illustrated in Fig.~\ref{fig:sim_setup}. For each position, we simulated a source uttering a \ac{WGN} signal, $10$s long for the labelled points and  a speech signal, $5$s long for the unlabelled points. The algorithm was tested on $n_T=200$ measurements of unknown sources from unknown locations with unique $5$s long speech signals. All the measurements were contaminated by additive \ac{WGN}. For each point, the \ac{CPSD} and the \ac{PSD} are estimated with Welch's method with $0.128$~s windows and $75\%$ overlap and are utilized for estimating the \ac{RTF} in~\eqref{eq:rtf_est} for $2048$ frequency bins. The \ac{RTF} vector consists of $D=286$ frequency bins corresponding to $0.2-2.5$kHz, in which most of the speech components are concentrated \BL{(for details please refer to~Appendix~\ref{sec:appA})}.   

\BLT{For the proposed method we used~\eqref{eq:recurs_est} to update the model according to the current test sample, i.e. for each test point the correlation is obtained by an average of $n_D+1$ points (the entire training set and the current test point).} \BLS{For comparison, we also examined the performance of two other algorithms which, although based on manifold considerations, heuristically fuse the data from the nodes.} Both algorithms rely on the manifold-based Gaussian process regression described in~\cite{laufer2016}. The first approach (`mean' in the graph) simply averages the estimates obtained by each single node separately. \BLS{The second algorithm (`Kernel-mult' in the graph) uses a Gaussian process with a covariance function that is given by the product of the individual kernels of the single nodes~\eqref{eq:gauss}. 
For a Gaussian kernel, using the product between the kernels of the different nodes is identical to using the \ac{aRTF} as an input to the kernel, i.e.
\begin{equation}
k(\mathbf{h}_i,\mathbf{h}_j)=k(\mathbf{h}^1_i,\mathbf{h}^1_j)\cdot k(\mathbf{h}^2_i,\mathbf{h}^2_j)\cdots k(\mathbf{h}^M_i,\mathbf{h}^M_j)
\end{equation}
since multiplying the kernels results in the summation of the squared distances, which equals the norm between the corresponding \acp{aRTF}. This means that the algorithm regards the \ac{aRTF} as a one long feature vector, and is indifferent to the fact that the measurements are aggregated by different nodes. In contrast, the proposed method individually refers to each node and its associated \ac{RTF}.}
As a baseline, we also compared the results with a modified version of the \ac{SRP-PHAT} algorithm~\cite{do2007real}. \BLT{Note that, opposed to the learning-based methods, the \ac{SRP-PHAT} algorithm requires the knowledge of the exact microphones' positions.}

\begin{figure}[ht!]
\centering
\subfigure[]{\includegraphics[trim=2cm 6cm 0.5cm 6.5cm, width=0.55\textwidth,height=0.3\textheight]{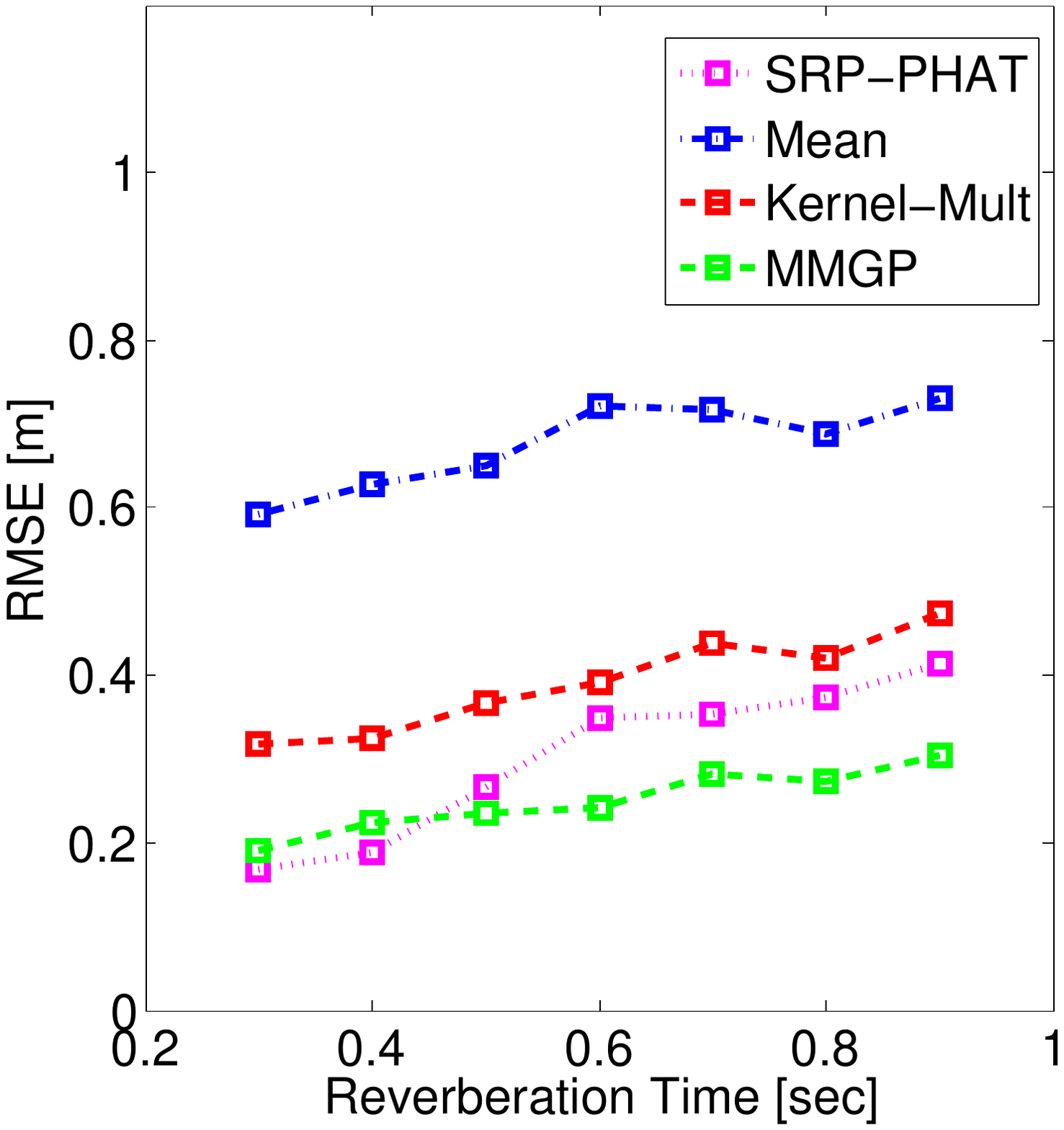}}
\subfigure[]{\includegraphics[trim=2cm 6cm 0.5cm 6.5cm, width=0.55\textwidth,height=0.3\textheight]{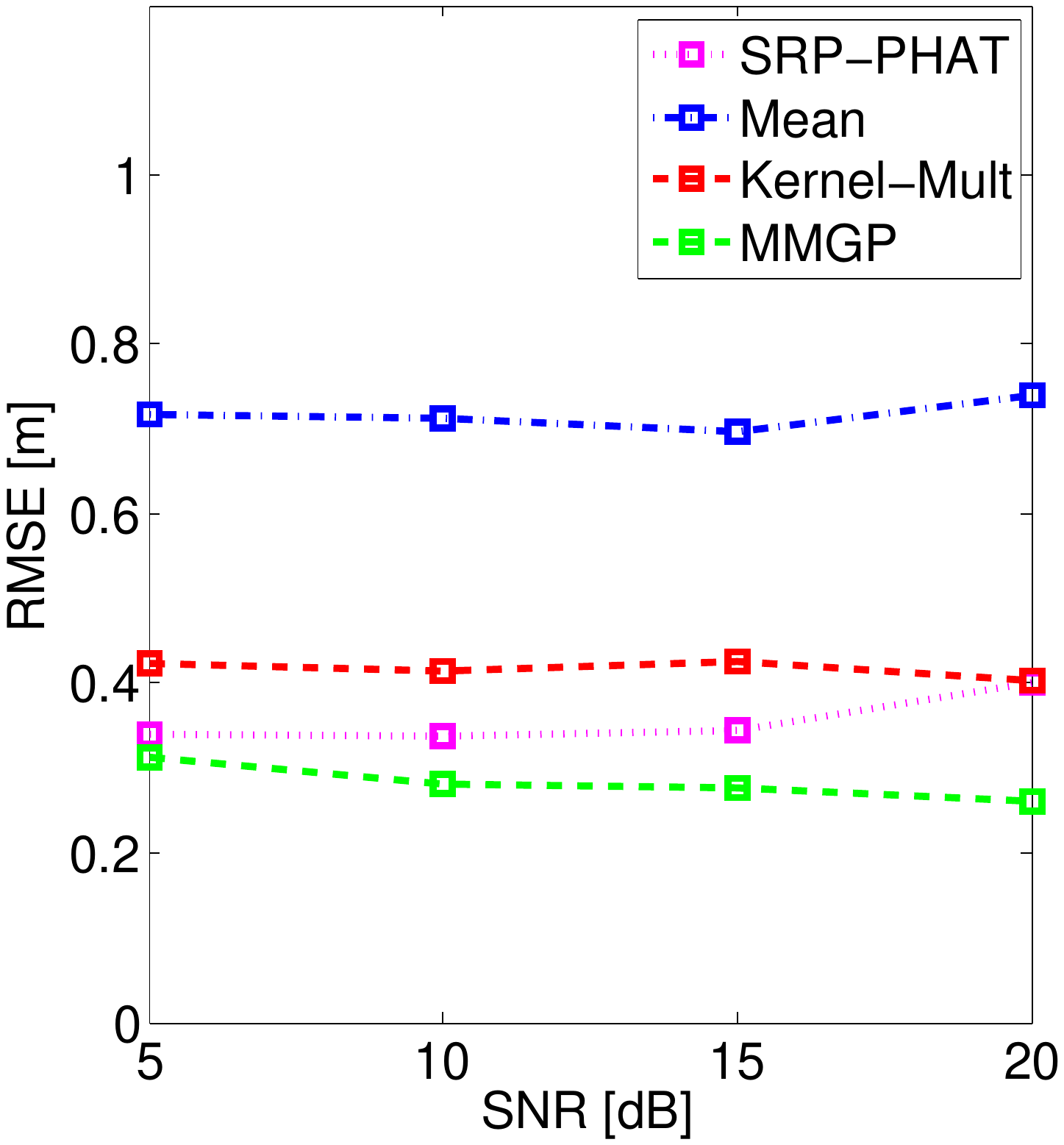}}
\centering
\caption{The \ac{RMSE} (a) for different reverberation times, and (b) for different noise levels.} 
\label{fig:res}
\end{figure}

The \acp{RMSE} attained by all four algorithms are compared in two scenarios. In the first scenario, different reverberation levels are examined while the \ac{SNR} is set to $20$dB. In the second scenario, the \ac{SNR} is varying while the reverberation time is set to $700$ms. In all scenarios, the training set is generated with a fixed \ac{SNR} of $20$dB. All the results are summarised in Fig.~\ref{fig:res}.  

It can be observed that the reverberation level has a direct influence on the performance, and all four algorithms exhibit degraded performance as reverberation increases. Regarding noise, it can be seen that the \ac{SNR} level does not have a clear impact on the performance. From the comparison between the algorithms it is indicated that the proposed method outperforms the other learning-based algorithms and obtains a significantly smaller error. The \ac{SRP-PHAT} performs better for lower reverberation levels, yet it is inferior for high reverberation levels. In addition, the proposed method obtains a smaller error compared to the \ac{SRP-PHAT} for all noise levels, in high reverberation conditions. 

\subsection{Real Recordings}
\begin{figure}[ht!]
\centering
\subfigure[]{\includegraphics[trim=1cm 7cm 1cm 6cm, width=0.45\textwidth,height=0.28\textheight]{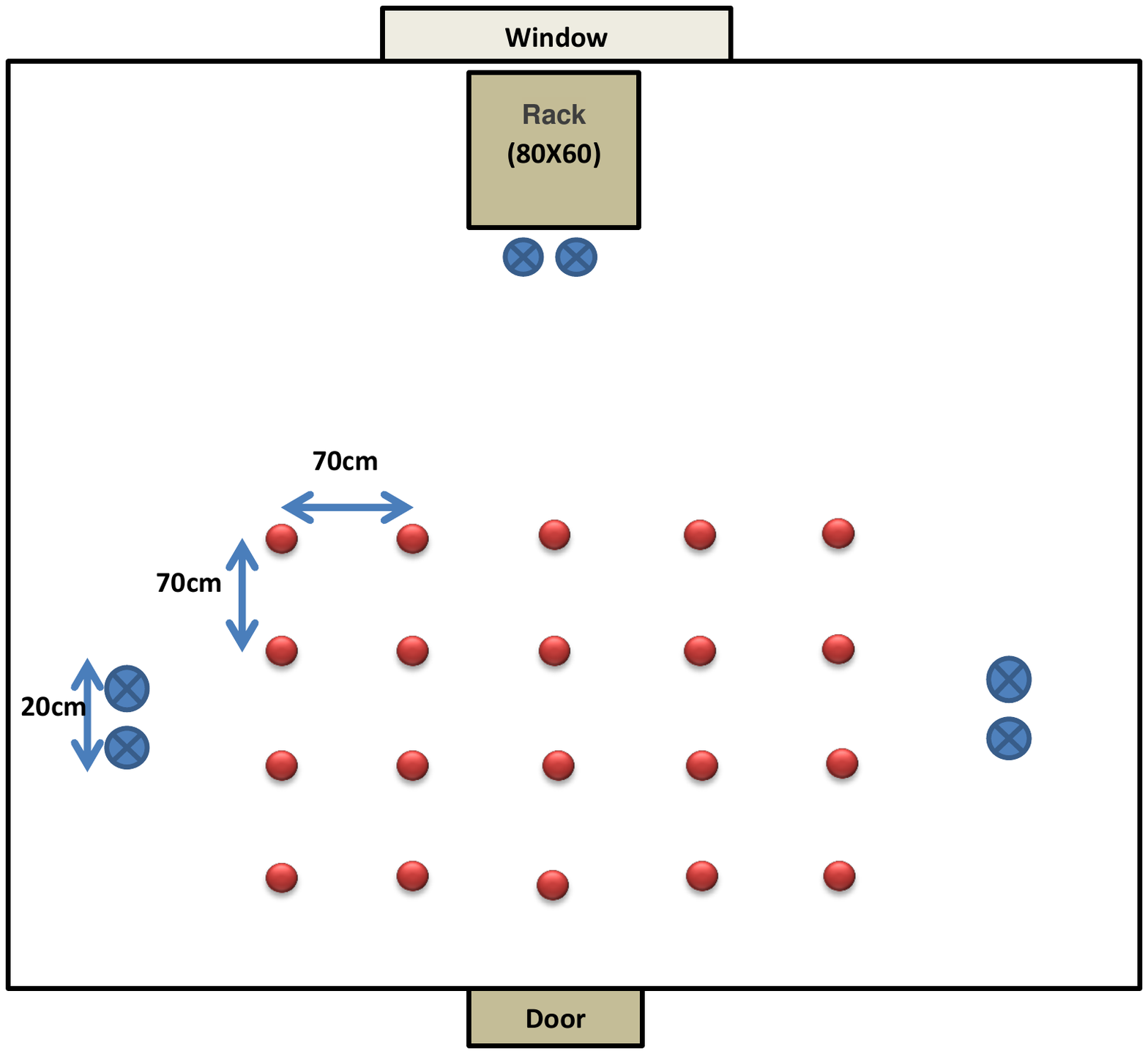}}
\subfigure[]{\includegraphics[trim=0.8cm 2.5cm 1cm 0.5cm width=0.25\textwidth,height=0.21\textheight]{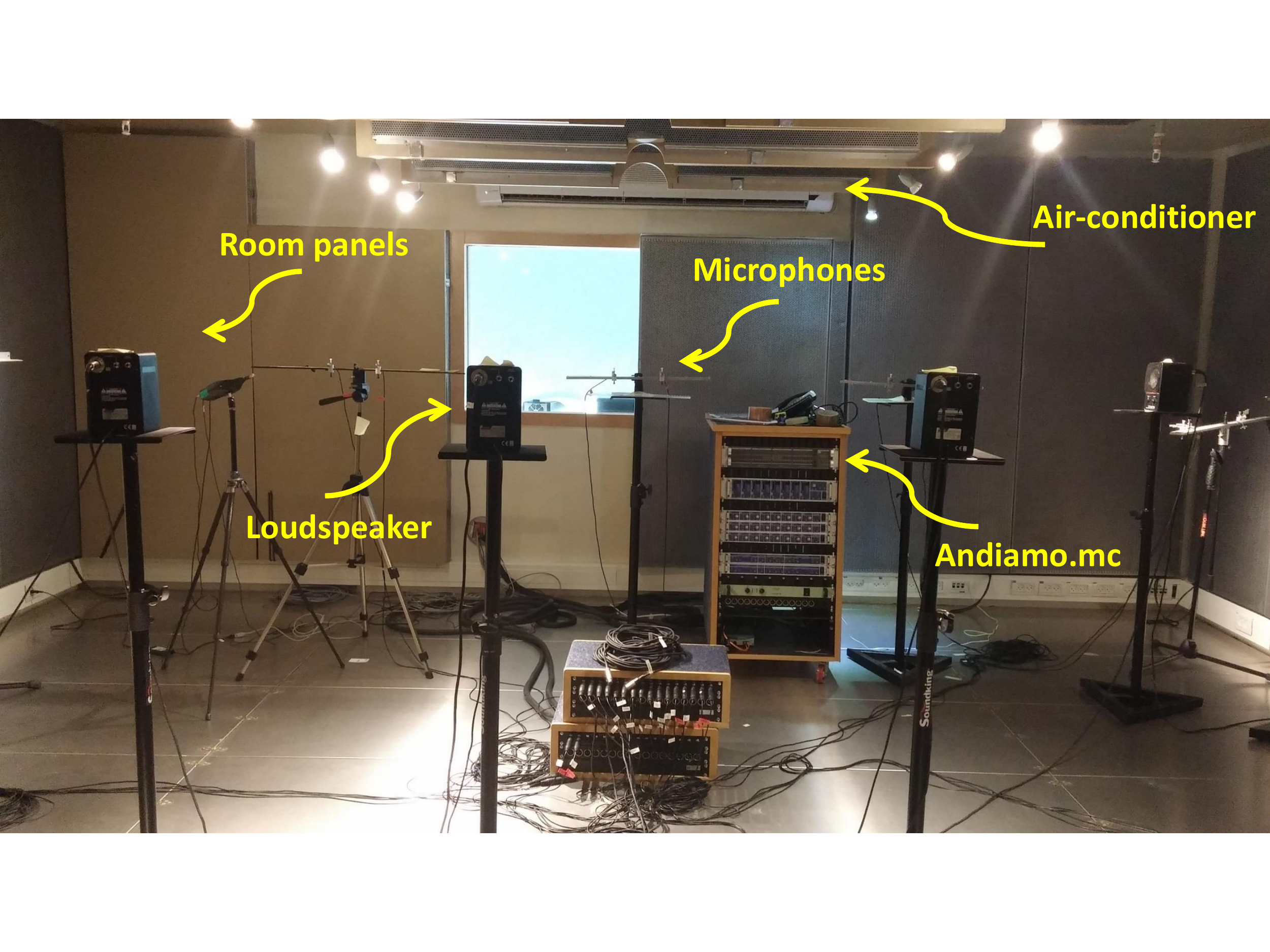}}
\centering
\caption{(a) The room layout: the microphone positions are marked by blue `x' marks, and the positions of the labelled samples are marked by red circles. (b) a photograph of the room.} 
\label{fig:lab}
\end{figure}

The algorithm performance was also tested using real recordings carried out in the speech and acoustic lab of Bar-Ilan University. This is a $6\times 6\times 2.4$m room controllable reverberation time, utilizing $60$ interchangeable panels covering the room facets. The measurement equipment consists of an RME Hammerfall HDSPe MADI sound-card and an Andiamo.mc (for Microphone pre-amplification and digitization (A/D)). As sources we used Fostex 6301BX loudspeakers which have a rather flat response in the frequency range $80$Hz-$13$kHz. The signals were measured by $6$ AKG type CK-32 omnidirectional microphones, which were placed in pairs with internal distance of $0.2$m. All the measurements were carried out with a sampling frequency of $48$kHz and a resolution of $24$-bits. The measured signals were then downsampled to $16$kHz. The reverberation level was set to $T_{60}=620$ms which was determined by changing the panels configuration. An illustration of the room layout is depicted in Fig.~\ref{fig:lab}(a) and a photograph of the room and the experimental setup is presented in Fig.~\ref{fig:lab}(b).

The source position is confined to a $2.8\times 2.1$m area located near the room entrance. In this region, we generated $n_L=20$ equally-spaced labelled samples with resolution of $0.7$m. Additional $n_U=50$ unlabelled measurements, were generated in this region in random positions. The algorithm performance was examined on $25$ test samples also generated in random positions, in the defined region. For generating the labelled samples a chirp signal, $30$s long, was used, while for generating both the unlabelled samples and the test samples we used $75$ different speech signals of both males and females, each $10$s long, drawn from the TIMIT database. The \ac{RTF} 
estimation was performed similarly to the way it was defined in the simulation part.

\begin{figure}[ht!]
\centering
\subfigure[]
{\includegraphics[trim=2cm 6cm 0.5cm 6.5cm, width=0.53\textwidth,height=0.32\textheight]{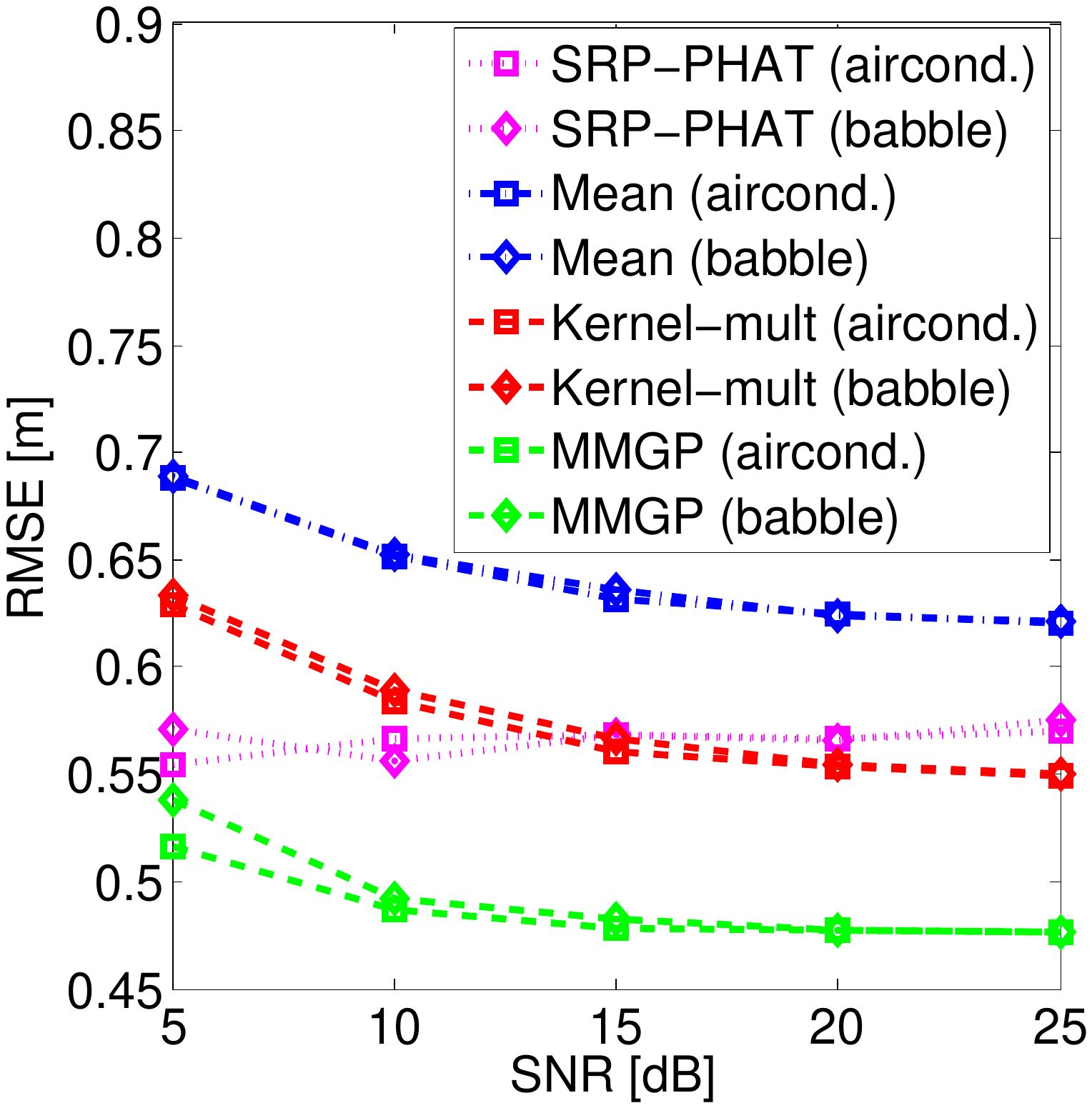}}
\centering
\caption{The \ac{RMSE} for different noise levels with two types of noise}
\label{fig:comp_snr}
\end{figure}

We examine two different types of noise sources: air-conditioner noise and babble noise, which is simultaneously played from $3$ loudspeakers located in the room. The \acp{RMSE} obtained for different \ac{SNR} levels when the reverberation is fixed to $T_{60}=620$ms, are depicted in Fig.~\ref{fig:comp_snr}(a). 
We observe that the proposed algorithm outperforms the other methods and obtains a smaller error for both noise types. It can also be observed that the results obtained based on the lab recordings exhibit the same trends as the results based on the simulated data.

\begin{figure}[ht!]
\centering
\includegraphics[trim=0.005cm 5cm 0.005cm 5cm, width=0.45\textwidth,height=0.3\textheight]{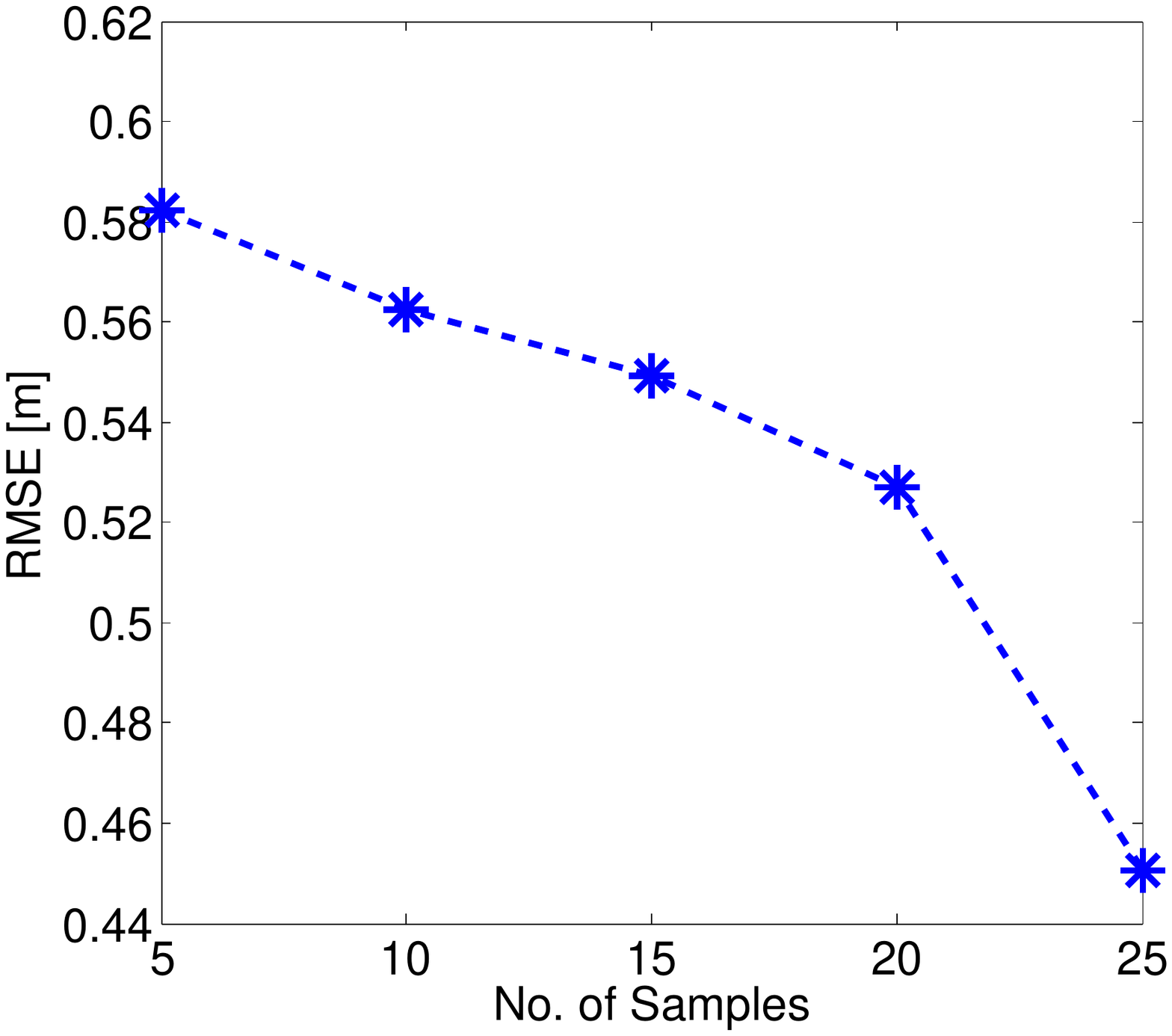}
\centering
\caption{\BLS{Demonstration of the recursive adaptation process:} in each step the current sample is used to update the covariance function of the process. The results are averaged over groups of $5$ samples.} 
\label{fig:recurs}
\end{figure}

\BLT{We also applied the recursive adaptation process presented in Section~\ref{sec:recurs}. The positions of the $25$ test samples are estimated sequentially where in each time step, the current sample is treated as an additional unlabelled sample and is used to update the covariance of the \ac{MMGP} according to~\eqref{eq:recurs_cov} and~\eqref{eq:recurs_est}. 
The samples in the test set are initially ordered  according to their physical adjacency, so that neighbouring samples are added in a sequential manner. \BLS{We use the same set of samples and repeat the sequential adaptation when applied to different orders of the samples in the set, by mixing the order of neighbouring samples.} In addition, we average the error for sets of $5$ consecutive time steps. Both averages are essential for the sake of generality to ensure that the results are neither tailored to a specific ordering of the samples in the set, nor reflect the quality of a particular sample. Figure~\ref{fig:recurs} depicts the average \ac{RMSE}. We observe a monotonic decrease in the error as more samples are added to the computation of the covariance function in a recursive manner. These results also emphasize the importance of the semi-supervised approach, i.e. the significant role that unlabelled samples have in the estimation process.}         
 
\begin{figure}[ht!]
\centering
\includegraphics[trim=0.5cm 6cm 0.5cm 6cm, width=0.5\textwidth,height=0.3\textheight]{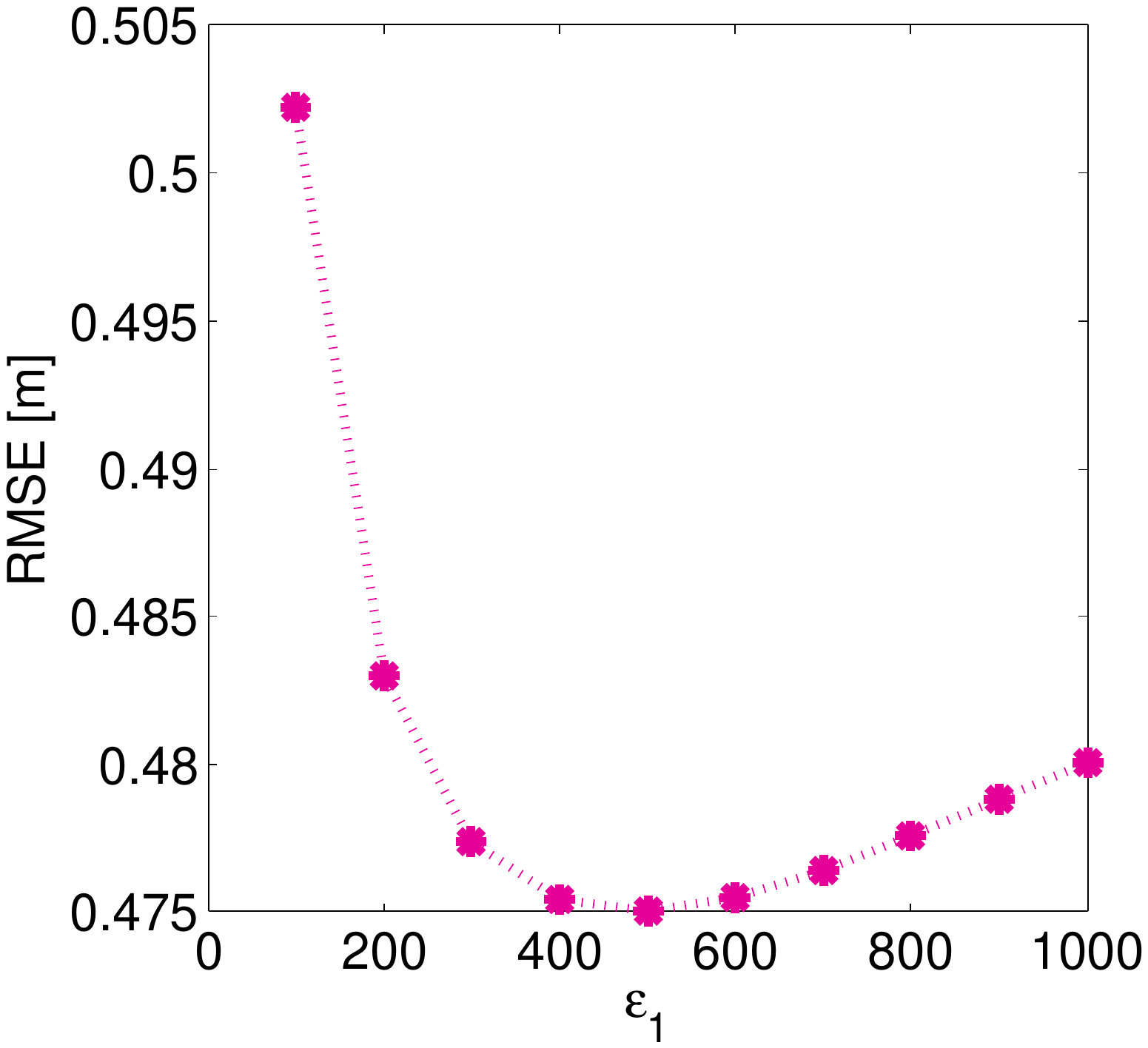}
\centering
\caption{The \ac{RMSE} obtained for different values of $\varepsilon_1$.} 
\label{fig:opt_eps}
\end{figure}

\BLS{Another examination was carried out to inspect the effectiveness of the parameter optimization through the \ac{ML} criterion of the labelled samples, as presented in Section~\ref{sec:parameters}. In Fig.~\ref{fig:opt_eps}, we present the error of the estimated test positions obtained for different values of $\varepsilon_1$ in the range between $100-1000$, while the other parameters remain fixed. It can be observed that the optimal value is around $500$. For comparison, we followed the proposed optimization using gradient decent starting from an initial value of $100$. We obtain that the optimal value for  $\varepsilon_1$ is $514$, which resembles the empirical value that optimized the performance on the test samples as implied by Fig.~\ref{fig:opt_eps}. This indicates that the parameter values, obtained through an optimization over the labelled samples, yields in practice plausible results for estimating the unknown positions of the test samples.}     

Finally, we investigated the effect of changes in the environmental conditions between the training and the test stages. \BLS{Training-based approaches are often criticized for being impractical, since identical conditions in both the training and the test phases cannot be guaranteed (e.g. door and windows may be opened or closed, people may move in the room etc.).} We examined two types of changes: the door of the room changed from closed (during training) to open (during test) and slight changes in the panel configuration (decreasing the room reverberation time by about $5\%$). We repeated the measurements of $20$ test samples in both scenarios (the training samples are left unchanged), and compared the results obtained under these conditions to the nominal results, where there is no change in the environmental conditions between the training set and the test set. This comparison is summarized in Table~\ref{tab:chang}, which presents the \acp{RMSE} in  all the defined scenarios. It can be seen that either opening the door or changing the panel configuration does not have a significant impact on the localization results of the proposed method, which indicates that the algorithm is robust to slight changes that are likely to occur in practical scenarios. Note that the results of the \ac{SRP-PHAT} algorithm are slightly improved under these changes due to the reduction in the reverberation level.  

\begin{table}[ht]
\centering
\begin{tabular}{|c|c|c|c|}
   \hline
   & & &\\ 
   & \textbf{Nominal} & \textbf{Door} & \textbf{Panel} \\ %\cline{1-4} 
   & & &  \\ \cline{1-4}   
   & & &  \\
   \textbf{MMGP}& $0.465$ & $0.493$& $0.506$ \\%\cline{1-4} 
   & & &  \\ \cline{1-4}    
   & & &  \\
  \textbf{\ac{SRP-PHAT}} & $0.540$ & $0.516$ & $0.531$\\ %\hline   
   & & &  \\ \cline{1-4}
\end{tabular}
\caption{Comparison between the \ac{RMSE} obtained in the case where the training and the test sets are generated exactly with the same conditions (first column) and when the test is generated under some environmental changes: open door (second column) or changes in the panel configuration (third column).}
\label{tab:chang}
\vspace{-0.4cm}
\end{table}

\section{Conclusions}
\label{sec:conclusions}
\BLT{In this paper, a novel mathematical approach was developed to fuse the information acquired in a multi-node scenario. This approach, when applied to source localization in ad hoc networks of distributed microphones, deviates from the common practice in the field since it is devised in a semi-supervised manner based on a data-driven model rather than on mathematically predefined relationships. A Gaussian process is used for modelling the unknown relation between the acoustic measurements and the corresponding source positions. The prerecorded training measurements provide useful information about the characteristics of the acoustic environment, and are used to define the covariance of the Gaussian process by averaging over both the different nodes and the different relations to other available acoustic samples. As for the practical aspect, the method produces satisfactory results in challenging adverse conditions including high reverberation and noise levels, with no need for microphone calibration (the algorithm is blind to their positions). The experimental results based on real lab recordings further emphasize the applicability of the algorithm and its ability to successfully locate the source in involved scenarios with possibly natural variations between the training and the test phases. Moreover, the gradual improvement in the performance, as demonstrated in the sequential application of the algorithm, verify the relevance of the information manifested in unlabelled training recordings to the localization task.} 

\begin{appendices}
%\appendix
\section{}
\label{sec:appA}
We consider the relative impulse response $h^m(n,\mathbf{p})$, which satisfies: $a^m_2(n,\mathbf{p})=h^m(n,\mathbf{p})*a^m_1(n,\mathbf{p})$. The \ac{AIR} is typically very long and complicated since it consists of the direct path between the source and the relevant microphone, and the various reflections from the different surfaces and objects in the enclosure. Thus, the relative impulse response  also has a complex high-dimensional nature. \BL{However, in a static environment where the acoustic conditions and the microphones position' are fixed, the only parameter that distinguishes between the different \acp{AIR} is the source position.} For convenience, we work in the frequency domain, and use the \acf{RTF} $H^m(k,\mathbf{p})$, which is the Fourier transform of the relative impulse response $h^m(n,\mathbf{p})$, where $k$ is the frequency index. Accordingly, the $m$th \ac{RTF} is given by the ratio between the two \acfp{ATF} of the two microphones in the $m$th pair, i.e. $H^m(k,\mathbf{p})=A^m_2(k,\mathbf{p})/A^m_1(k,\mathbf{p})$, where $A^m_i(k,\mathbf{p})$ is the \ac{ATF} of the respective \ac{AIR} $a^m_i(n,\mathbf{p})$. \BL{Assuming uncorrelated noise,} the $m$th \ac{RTF} can be computed using the \ac{PSD} and \ac{CPSD} of the measured signals and the noise at the $m$th pair:
\begin{multline}
H^m(k,\mathbf{p}) = \frac{S^m_{y_2y_1}(k,\mathbf{p})}{S^m_{y_1y_1}(k,\mathbf{p})-S^m_{u_1u_1}(k)} =\\\frac{S_{ss}(k)A^m_2(k,\mathbf{p})A_1^{m*}(k,\mathbf{p})}{S_{ss}(k)|A^m_1(k,\mathbf{p})|^2}
=\frac{A^m_2(k,\mathbf{p})}{A^m_1(k,\mathbf{p})} 
\label{eq:rtf}
\end{multline}        
where $S^m_{y_2y_1}(k,\mathbf{p})$ is the \ac{CPSD} between $y^m_1(n)$ and $y^m_2(n)$, $S^m_{y_1y_1}(k,\mathbf{p})$ is the \ac{PSD} of $y^m_1(n)$, $S^m_{u_1u_1}(k)$ is the \ac{PSD} of the noise $u^m_1(n)$ in the first microphone, and $S_{ss}(k)$ is the \ac{PSD} of the source $s(n)$. We use a biased estimator of the \ac{RTF}, neglecting the noise \ac{PSD} in the denominator of~\eqref{eq:rtf}:
\begin{equation}
\hat{H}^m(k,\mathbf{p}) \equiv \frac{\hat{S}^m_{y_2y_1}(k,\mathbf{p})}{\hat{S}^m_{y_1y_1}(k,\mathbf{p})} .
\label{eq:rtf_est}
\end{equation}
where $\hat{S}^m_{y_2y_1}(k,\mathbf{p})$ and $\hat{S}^m_{y_1y_1}(k,\mathbf{p})$ are estimated based on the measured signals. 
Let $\mathbf{h}^m(\mathbf{p})=[\hat{H}^m(k_1,\mathbf{p}), \ldots ,\hat{H}^m(k_{D},\mathbf{p})]^T$, be a concatenation of \ac{RTF} estimates of the $m$th node in $D$ frequency bins. Due to the symmetry of the Fourier transform for real valued functions, only the first half of the transform is considered. In addition, we consider only those frequency bins where the speech components are most likely to be present, to avoid poor estimates of~\eqref{eq:rtf_est} in frequencies where the speech components are absent. For the sake of clarity, the position index is omitted throughout the paper.
\end{appendices}
% -------------------------------------------------------------------------
% Either list references using the bibliography style file IEEEtran.bst
\balance
\bibliographystyle{IEEEtran}

%\bibliography{papers}

\begin{thebibliography}{10}
\providecommand{\url}[1]{#1}
\csname url@samestyle\endcsname
\providecommand{\newblock}{\relax}
\providecommand{\bibinfo}[2]{#2}
\providecommand{\BIBentrySTDinterwordspacing}{\spaceskip=0pt\relax}
\providecommand{\BIBentryALTinterwordstretchfactor}{4}
\providecommand{\BIBentryALTinterwordspacing}{\spaceskip=\fontdimen2\font plus
\BIBentryALTinterwordstretchfactor\fontdimen3\font minus
  \fontdimen4\font\relax}
\providecommand{\BIBforeignlanguage}[2]{{%
\expandafter\ifx\csname l@#1\endcsname\relax
\typeout{** WARNING: IEEEtran.bst: No hyphenation pattern has been}%
\typeout{** loaded for the language `#1'. Using the pattern for}%
\typeout{** the default language instead.}%
\else
\language=\csname l@#1\endcsname
\fi
#2}}
\providecommand{\BIBdecl}{\relax}
\BIBdecl

\bibitem{huang2000passive}
Y.~Huang, J.~Benesty, and G.~W. Elko, ``Passive acoustic source localization
  for video camera steering,'' in \emph{IEEE International Conference on
  Acoustics, Speech, and Signal Processing (ICASSP)}, vol.~2, 2000, pp.
  909--912.

\bibitem{mandel2010model}
M.~I. Mandel, R.~J. Weiss, and D.~P. Ellis, ``Model-based
  expectation-maximization source separation and localization,'' \emph{IEEE
  Transactions on Audio, Speech, and Language Processing}, vol.~18, no.~2, pp.
  382--394, 2010.

\bibitem{nakadai2002real}
K.~Nakadai, H.~G. Okuno, H.~Kitano \emph{et~al.}, ``Real-time sound source
  localization and separation for robot audition.'' in \emph{INTERSPEECH},
  2002.

\bibitem{valin2003robust}
J.-M. Valin, F.~Michaud, J.~Rouat, and D.~L{\'e}tourneau, ``Robust sound source
  localization using a microphone array on a mobile robot,'' in
  \emph{Proceedings of IEEE/RSJ International Conference on Intelligent Robots
  and Systems (IROS)}, vol.~2.\hskip 1em plus 0.5em minus 0.4em\relax IEEE,
  2003, pp. 1228--1233.

\bibitem{hornstein2006sound}
J.~Hornstein, M.~Lopes, J.~S. Victor, and F.~Lacerda, ``Sound localization for
  humanoid robots-building audio-motor maps based on the hrtf,'' in
  \emph{Proceedings of IEEE/RSJ International Conference on Intelligent Robots
  and Systems (IROS)}, 2006, pp. 1170--1176.

\bibitem{yao2002maximum}
K.~Yao, J.~C. Chen, and R.~E. Hudson, ``Maximum-likelihood acoustic source
  localization: experimental results,'' in \emph{IEEE International Conference
  on Acoustics, Speech, and Signal Processing (ICASSP)}, vol.~3, 2002, pp.
  2949--2952.

\bibitem{schmidt1986multiple}
R.~O. Schmidt, ``Multiple emitter location and signal parameter estimation,''
  \emph{IEEE Transactions on Antennas and Propagation}, vol.~34, no.~3, pp.
  276--280, 1986.

\bibitem{roy1989esprit}
R.~Roy and T.~Kailath, ``{ESPRIT}-estimation of signal parameters via
  rotational invariance techniques,'' \emph{IEEE Transactions on Acoustics,
  Speech and Signal Processing}, vol.~37, no.~7, pp. 984--995, 1989.

\bibitem{Knapp1976}
C.~Knapp and G.~Carter, ``The generalized correlation method for estimation of
  time delay,'' \emph{IEEE Transactions on Acoustic, Speech and Signal
  Processing}, vol.~24, no.~4, pp. 320--327, Aug. 1976.

\bibitem{brandstein1997robust}
M.~S. Brandstein and H.~F. Silverman, ``A robust method for speech signal
  time-delay estimation in reverberant rooms,'' in \emph{IEEE International
  Conference on Acoustics, Speech and Signal Processing (ICASSP)}, vol.~1,
  1997, pp. 375--378.

\bibitem{stephenne1997new}
A.~St{\'e}phenne and B.~Champagne, ``A new cepstral prefiltering technique for
  estimating time delay under reverberant conditions,'' \emph{Signal
  Processing}, vol.~59, no.~3, pp. 253--266, 1997.

\bibitem{rui2004time}
Y.~Rui and D.~Florencio, ``Time delay estimation in the presence of correlated
  noise and reverberation,'' in \emph{IEEE International Conference on
  Acoustics, Speech, and Signal Processing (ICASSP)}, vol.~2, 2004, pp.
  133--136.

\bibitem{dvorkind2005time}
T.~Dvorkind and S.~Gannot, ``Time difference of arrival estimation of speech
  source in a noisy and reverberant environment,'' \emph{Signal Processing},
  vol.~85, no.~1, pp. 177--204, Jan. 2005.

\bibitem{scheuing2008disambiguation}
J.~Scheuing and B.~Yang, ``Disambiguation of tdoa estimation for multiple
  sources in reverberant environments,'' \emph{IEEE Transactions on Audio,
  Speech, and Language Processing}, vol.~16, no.~8, pp. 1479--1489, 2008.

\bibitem{benesty2000adaptive}
J.~Benesty, ``Adaptive eigenvalue decomposition algorithm for passive acoustic
  source localization,'' \emph{The Journal of the Acoustical Society of
  America}, vol. 107, no.~1, pp. 384--391, 2000.

\bibitem{doclo2003robust}
S.~Doclo and M.~Moonen, ``Robust adaptive time delay estimation for speaker
  localization in noisy and reverberant acoustic environments,'' \emph{EURASIP
  Journal on Applied Signal Processing}, vol. 2003, pp. 1110--1124, 2003.

\bibitem{dibiase2001robust}
J.~H. DiBiase, H.~F. Silverman, and M.~S. Brandstein, ``Robust localization in
  reverberant rooms,'' in \emph{Microphone Arrays}.\hskip 1em plus 0.5em minus
  0.4em\relax Springer, 2001, pp. 157--180.

\bibitem{deleforge20122d}
A.~Deleforge and R.~Horaud, ``{2D} sound-source localization on the binaural
  manifold,'' in \emph{IEEE International Workshop on Machine Learning for
  Signal Processing (MLSP)}, Santander, Spain, Sep. 2012.

\bibitem{deleforge2013variational}
A.~Deleforge, F.~Forbes, and R.~Horaud, ``Variational {EM} for binaural
  sound-source separation and localization,'' in \emph{IEEE International
  Conference on Acoustics, Speech and Signal Processing (ICASSP)}, 2013, pp.
  76--80.

\bibitem{deleforge2015acoustic}
------, ``Acoustic space learning for sound-source separation and localization
  on binaural manifolds,'' \emph{International journal of neural systems},
  vol.~25, no.~1, 2015.

\bibitem{may2011probabilistic}
T.~May, S.~Van De~Par, and A.~Kohlrausch, ``A probabilistic model for robust
  localization based on a binaural auditory front-end,'' \emph{IEEE
  Transactions on Audio, Speech, and Language Processing}, vol.~19, no.~1, pp.
  1--13, 2011.

\bibitem{Wu2016spatial}
X.~Wu, D.~S. Talagalaz, and T.~D. Abhayapalay, ``Spatial feature learning for
  robust binaural sound source localization using a composite feature vector,''
  in \emph{IEEE International Conference on Acoustics, Speech and Signal
  Processing (ICASSP)}, Shanghai, China, Mar. 2016.

\bibitem{xiong2015}
X.~Xiao, S.~Zhao, X.~Zhong, D.~L. Jones, E.~S. Chng, and H.~Li, ``A
  learning-based approach to direction of arrival estimation in noisy and
  reverberant environments,'' in \emph{IEEE International Conference on
  Acoustics, Speech and Signal Processing (ICASSP)}, 2015, pp. 76--80.

\bibitem{xiong2016spatial}
X.~Xiao, S.~Zhao, T.~N.~T. Nguyen, D.~L. Jones, E.~S. Chng, and H.~Li,
  ``Spatial feature learning for robust binaural sound source localization
  using a composite feature vector,'' in \emph{IEEE International Conference on
  Acoustics, Speech and Signal Processing (ICASSP)}, Shanghai, China, Mar.
  2016.

\bibitem{kitic2014hearing}
S.~Kitic, N.~Bertin, and R.~Gribonval, ``Hearing behind walls: localizing
  sources in the room next door with cosparsity,'' in \emph{IEEE International
  Conference on Acoustics, Speech and Signal Processing (ICASSP)}, 2014, pp.
  3087--3091.

\bibitem{bertin2016joint}
N.~Bertin, S.~Kitic, and R.~Gribonval, ``Joint estimation of sound source
  location and boundary impedance with physics-driven cosparse
  regularization,'' in \emph{IEEE International Conference on Acoustics, Speech
  and Signal Processing (ICASSP)}, Shanghai, China, Mar. 2016.

\bibitem{talmon2012parametrization}
R.~Talmon, D.~Kushnir, R.~Coifman, I.~Cohen, and S.~Gannot, ``Parametrization
  of linear systems using diffusion kernels,'' \emph{IEEE Transactions on
  Signal Processing}, vol.~60, no.~3, pp. 1159--1173, Mar. 2012.

\bibitem{Talmon2011}
R.~Talmon, I.~Cohen, and S.~Gannot, ``Supervised source localization using
  diffusion kernels,'' in \emph{IEEE Workshop on Applications of Signal
  Processing to Audio and Acoustics (WASPAA)}, 2011, pp. 245--248.

\bibitem{laufer2013relative}
B.~Laufer-Goldshtein, R.~Talmon, and S.~Gannot, ``Relative transfer function
  modeling for supervised source localization,'' in \emph{IEEE Workshop on
  Applications of Signal Processing to Audio and Acoustics (WASPAA)}, New
  Paltz, NY, USA, Oct. 2013.

\bibitem{laufer2016semi}
------, ``Semi-supervised sound source localization based on manifold
  regularization,'' \emph{IEEE Transactions on Audio, Speech, and Language
  Processing}, vol.~24, no.~8, pp. 1393--1407, 2016.

\bibitem{laufer2016}
------, ``Manifold-based bayesian inference for semi-supervised source
  localization,'' in \emph{IEEE International Conference on Acoustics, Speech
  and Signal Processing (ICASSP)}, Shanghai, China, Mar. 2016.

\bibitem{sindhwani2005beyond}
V.~Sindhwani, P.~Niyogi, and M.~Belkin, ``Beyond the point cloud: from
  transductive to semi-supervised learning,'' in \emph{Proceedings of the 22nd
  international conference on Machine learning}.\hskip 1em plus 0.5em minus
  0.4em\relax ACM, 2005, pp. 824--831.

\bibitem{sindhwani2007semi}
V.~Sindhwani, W.~Chu, and S.~S. Keerthi, ``Semi-supervised gaussian process
  classifiers.'' in \emph{IJCAI}, 2007, pp. 1059--1064.

\bibitem{Gannot2001}
S.~Gannot, D.~Burshtein, and E.~Weinstein, ``Signal enhancement using
  beamforming and nonstationarity with applications to speech,'' \emph{IEEE
  Transactions on Signal Processing}, vol.~49, no.~8, pp. 1614 --1626, Aug.
  2001.

\bibitem{markovich2009multichannel}
S.~Markovich, S.~Gannot, and I.~Cohen, ``Multichannel eigenspace beamforming in
  a reverberant noisy environment with multiple interfering speech signals,''
  \emph{IEEE Transactions on Audio, Speech, and Language Processing}, vol.~17,
  no.~6, pp. 1071--1086, 2009.

\bibitem{laufer2015}
B.~Laufer-Goldshtein, R.~Talmon, and S.~Gannot, ``Study on manifolds of
  acoustic responses,'' in \emph{Interntional Conference on Latent Variable
  Analysis and Signal Seperation (LVA/ICA)}, Liberec, Czech Republic, Aug.
  2015.

\bibitem{rasmussen2006gaussian}
C.~E. Rasmussen and C.~K.~I. Williams, \emph{Gaussian processes for machine
  learning}.\hskip 1em plus 0.5em minus 0.4em\relax MIT Press, 2006.

\bibitem{kushnir2012anisotropic}
D.~Kushnir, A.~Haddad, and R.~R. Coifman, ``Anisotropic diffusion on
  sub-manifolds with application to earth structure classification,''
  \emph{Applied and Computational Harmonic Analysis}, vol.~32, no.~2, pp.
  280--294, 2012.

\bibitem{haddad2014texture}
A.~Haddad, D.~Kushnir, and R.~R. Coifman, ``Texture separation via a reference
  set,'' \emph{Applied and Computational Harmonic Analysis}, vol.~36, no.~2,
  pp. 335--347, 2014.

\bibitem{Coifman2006}
R.~Coifman and S.~Lafon, ``Diffusion maps,'' \emph{Appl. Comput. Harmon.
  Anal.}, vol.~21, pp. 5--30, Jul. 2006.

\bibitem{lederman2015learning}
R.~R. Lederman and R.~Talmon, ``Learning the geometry of common latent
  variables using alternating-diffusion,'' \emph{Applied and Computational
  Harmonic Analysis}, 2015.

\bibitem{woodbury1950inverting}
M.~A. Woodbury, ``Inverting modified matrices,'' \emph{Memorandum report},
  vol.~42, p. 106, 1950.

\bibitem{Habets2006}
E.~A.~P. Habets, ``Room impulse response ({RIR}) generator,''
  https://www.audiolabs-erlangen.de/fau/professor/habets/software/rir-generator,
  Jul. 2006.

\bibitem{Image79}
J.~Allen and D.~Berkley, ``Image method for efficiently simulating small-room
  acoustics,'' \emph{J. Acoustical Society of America}, vol.~65, no.~4, pp.
  943--950, Apr. 1979.

\bibitem{do2007real}
H.~Do, H.~F. Silverman, and Y.~Yu, ``A real-time srp-phat source location
  implementation using stochastic region contraction (src) on a large-aperture
  microphone array,'' in \emph{IEEE International Conference on Acoustics,
  Speech and Signal Processing (ICASSP)}, vol.~1, 2007, pp. 121--124.

\end{thebibliography}

\end{document}